\documentclass[journal=jctcce,manuscript=article]{achemso}

\usepackage[version=3]{mhchem} % Formula subscripts using \ce{}
\usepackage{braket}
\usepackage{mathtools}
\usepackage{algorithm}
\usepackage{algorithmic}
\usepackage{geometry}
\usepackage{amsmath}
\usepackage{amsthm}
\usepackage{graphicx}
\usepackage{fancyvrb}
\usepackage[dvipsnames]{xcolor}
\author{Millan F. Welman}
\affiliation[Princeton University]
{Department of Chemistry, Princeton University, Princeton, NJ 08544}
\author{Tao E. Li}
\affiliation[University of Delaware]
{Department of Physics and Astronomy, University of Delaware, Newark, DE 19716}
\author{Sharon Hammes-Schiffer}
\email{shs566@princeton.edu}

\affiliation[Princeton University]
{Department of Chemistry, Princeton University, Princeton, NJ 08544}

\title
  {Light-Matter Entanglement in Real--Time Nuclear-Electronic Orbital Polariton Dynamics}

\keywords{American Chemical Society, \LaTeX}

\begin{document}
%%%%%%%%%%%%%%%%%%%%%%%%%%%%%%%%%%%%%%%%%%%%%%%%%%%%%%%%%%%%%%%%%%%%%
%% The "tocentry" environment can be used to create an entry for the
%% graphical table of contents. It is given here as some journals
%% require that it is printed as part of the abstract page. It will
%% be automatically moved as appropriate.

 \begin{tocentry}
        \centering\includegraphics[height=3.25cm, keepaspectratio]{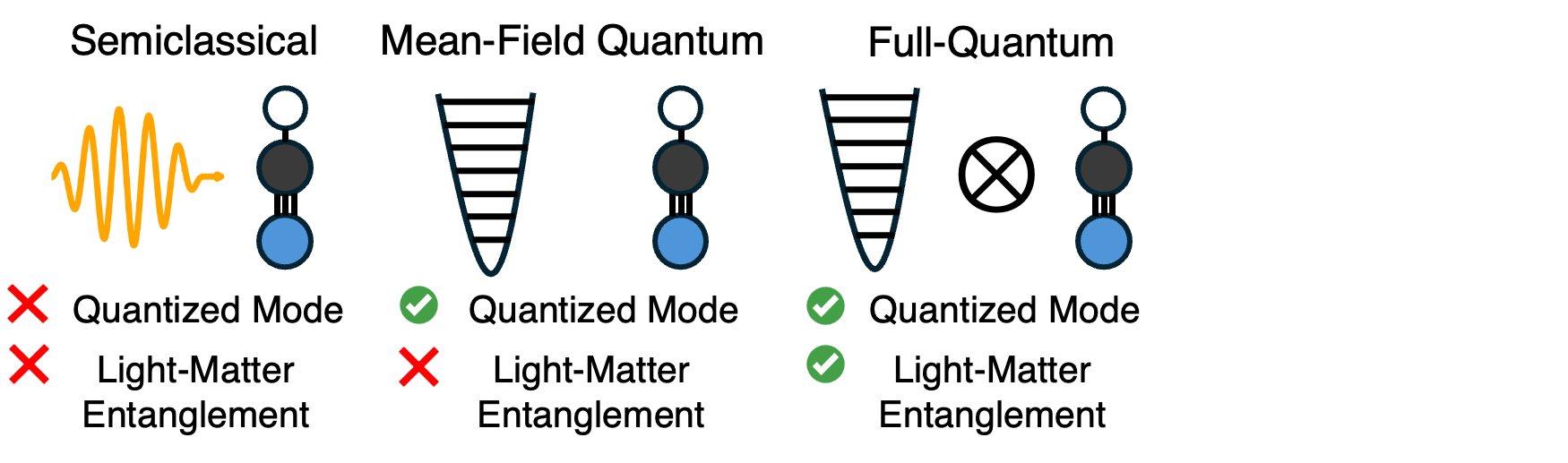}
	\end{tocentry}
%%%%%%%%%%%%%%%%%%%%%%%%%%%%%%%%%%%%%%%%%%%%%%%%%%%%%%%%%%%%%%%%%%%%%

%%%%%%%%%%%%%%%%%%%%%%%%%%%%%%%%%%%%%%%%%%%%%%%%%%%%%%%%%%%%%%%%%%%%%
%% The abstract environment will automatically gobble the contents
%% if an abstract is not used by the target journal.
%%%%%%%%%%%%%%%%%%%%%%%%%%%%%%%%%%%%%%%%%%%%%%%%%%%%%%%%%%%%%%%%%%%%%
\begin{abstract}
Molecular polaritons are hybrid light-matter states that enable the exploration of potential cavity-modified chemistry. The development of dynamical, first-principles approaches for simulating molecular polaritons is important for understanding their origins and properties. Herein, we present a hierarchy of first-principles methods to simulate the real-time dynamics of molecular polaritons in the strong coupling regime. These methods are based on real-time time-dependent density functional theory (RT-TDDFT) and the corresponding real-time nuclear-electronic orbital (RT-NEO) approach, in which specified nuclei are treated quantum mechanically on the same level as the electrons. The hierarchy spans semiclassical, mean-field-quantum, and full-quantum approaches to simulate polariton dynamics under both electronic strong coupling and vibrational strong coupling. In the semiclassical approaches, the cavity mode is treated classically, whereas in the full-quantum approaches, the cavity mode is treated quantum mechanically with propagation of a joint molecule-mode density matrix. The semiclassical and full-quantum approaches produce virtually identical Rabi splittings and polariton peak locations for the systems studied. However, the full-quantum approaches allow exploration of molecule-mode quantum entanglement in the real-time dynamics. Although the degree of light-matter entanglement is relatively small in the systems considered, the oscillations of the von Neumann entropy reveal an entanglement Rabi splitting that differs from the Rabi splitting computed from the time-dependent dipole moment. These results suggest that a classical treatment of the cavity mode may provide an excellent description of polariton dynamics for macroscopic observables such as the Rabi splitting, but novel physics may be detectable by considering molecule-mode entanglement.
\end{abstract}

%%%%%%%%%%%%%%%%%%%%%%%%%%%%%%%%%%%%%%%%%%%%%%%%%%%%%%%%%%%%%%%%%%%%%
%% Start the main part of the manuscript here.
%%%%%%%%%%%%%%%%%%%%%%%%%%%%%%%%%%%%%%%%%%%%%%%%%%%%%%%%%%%%%%%%%%%%%
\section{Introduction}
Molecular polaritons are hybrid light-matter states that form due to strong coupling between molecules and electromagnetic modes \cite{mandal_theoretical_2023}. The perturbation of a molecular system by an electromagnetic field creates a time-dependent induced molecular dipole moment. If the timescale of decoherence for the field is comparable to or greater than the timescale of oscillation of the induced dipole moment, a polariton will result from the coupling between the induced dipole and the incident field \cite{xiong_molecular_2023}. Coupling of the field to a transition dipole between two electronic states is referred to as the electronic strong coupling (ESC) regime \cite{deng_exciton-polariton_2010, keeling_boseeinstein_2020}, whereas coupling of the field to a transition dipole between two vibrational states is referred to as the vibrational strong coupling (VSC) regime \cite{james_p_long_coherent_2015, george_liquid-phase_2015, wright_rovibrational_2023}. These regimes can be realized experimentally in a variety of contexts ranging from optical Fabry-P\'{e}rot cavities \cite{xiang_molecular_2024}, where a field is trapped with a molecular system between a pair of mirrors in an enclosed volume, to plasmonic nanocavities \cite{pelton_strong_2019}, where a plasmonic mode is coupled to a small number of molecules. The last decade has seen extensive investigation into the possibility of cavity-modified chemistry in the strong-coupling regime, with previous experimental work reporting altered reaction rates \cite{ahn_modification_2023, thomas_groundstate_2016, lather_cavity_2019}, enhancements in energy and electron transfer as well as exciton transport \cite{zhong_energy_2017, xiang_intermolecular_2020, garcia-vidal_manipulating_2021, georgiou_ultralongrange_2021, kertzscher_tunable_2024, lidzey_strong_1998, feist_extraordinary_2015, schachenmayer_cavity-enhanced_2015, coles_polariton-mediated_2014, kasprzak_boseeinstein_2006}, and modifications in branching ratios \cite{thomas_tilting_2019}. \par

This abundance of experimental data has motivated the development of a broad array of promising methods to simulate and attempt to evaluate cavity-modified chemistry. Treatments based on the simple Jaynes-Cummings\cite{jaynes_comparison_1963} and Tavis-Cummings\cite{tavis_exact_1968, tavis_approximate_1969} two-level models from quantum optics have been extended over the past decade to incorporate greater detail in the treatment of molecular structure, including anharmonicity \cite{triana_shape_2020}, coupling to multiple cavity modes \cite{hoffmann_effect_2020, tichauer_multi-scale_2021}, and solvent interactions \cite{hoi_ling_luk_multiscale_2017}. Quantum-electrodynamical (QED) electronic structure theory methods \cite{christian_schafer_ab_2018, deprince_cavity-modulated_2021, mallory_reduced-density-matrix-based_2022, vu_cavity_2024, el_moutaoukal_strong_2025, bauer_perturbation_2023, mctague_non-hermitian_2022, pavosevic_wavefunction_2022, cui_variational_2024} such as QEDFT (quantum-electrodynamical density functional theory) \cite{flick_lightmatter_2019, ruggenthaler_quantum-electrodynamical_2014, flick_kohnsham_2015, schafer_shining_2022, flick_ab_2020, konecny_relativistic_2025}, which can be used to compute polaritonic potential energy surfaces, represent another valuable advance. \par

Although these theoretical methods have provided many useful insights, the development of additional approaches is desirable to investigate cavity-mediated chemistry. For this purpose, the method should provide a first-principles treatment of as many degrees of freedom as possible and describe molecular vibrations as well as electronic structure in their full complexity. Ideally, such theoretical methods would be scalable from a single molecule interacting with a single cavity mode to many molecules interacting with many cavity modes to allow a first-principles simulation of collective effects, which are believed to play a key role in cavity-mediated chemistry \cite{xiang_molecular_2024, del_pino_signatures_2015, campos-gonzalez-angulo_swinging_2023}. Dynamical approaches are desirable to provide a time-resolved description of interactions between molecular dipoles and cavity modes that give rise to the formation of polaritons. \par 

The dynamical description of polariton formation provides the option to treat the cavity modes strongly coupled to the molecular system either classically or quantum-mechanically. Both choices are viable for simulating molecular polaritons because the dipole-field coupling that creates the dynamical feedback underlying polariton formation occurs in both classical and quantum electrodynamics. A classical treatment of the field reduces the computational cost of a simulation, which permits the study of larger molecular systems and provides a route to the scalability required to describe the collective regime. Along these lines, the cavity molecular dynamics (CavMD) method \cite{li_cavity_2020} treats cavity modes as classical harmonic oscillators coupled to classical nuclear degrees of freedom propagated on an electronic ground-state potential energy surface, which may be computed with a molecular mechanical force field, electronic structure methods, or mixed quantum mechanical/molecular mechanical (QM/MM) methods \cite{li_qmmm_2023}. By treating all cavity and nuclear degrees of freedom classically, this method provides a computationally tractable approach for first-principles dynamical simulations of collective effects under VSC, where an ensemble of molecules is coupled to the cavity modes. \par
A classical treatment of the cavity modes was also recently combined with the real-time nuclear-electronic orbital TDDFT (RT-NEO-TDDFT) method \cite{li_semiclassical_2022} to create the semiclassical RT-NEO (sc-RT-NEO) approach. The RT-NEO method \cite{zhao_real-time_2020} extends conventional electronic RT-TDDFT to the NEO framework. In the NEO framework \cite{webb_multiconfigurational_2002, hammes-schiffer_nuclearelectronic_2021}, selected light nuclei, usually protons, are treated quantum mechanically on the same footing as the electronic degrees of freedom. The RT-NEO method self-consistently propagates real-time nonequilibrium electronic and quantum nuclear dynamics in a manner that directly accounts for nuclear quantum effects such as proton delocalization, zero-point energy, and tunneling. The sc-RT-NEO method incorporates the self-consistent propagation of classical cavity modes coupled to the molecular electronic and/or nuclear degrees of freedom, thereby enabling first-principles simulations of molecular polaritons in the ESC and VSC regimes. \par 

The classical treatment of cavity modes in CavMD and sc-RT-NEO has precedence in the literature. Semiclassical descriptions of light-matter interactions, where a molecular system treated quantum mechanically is coupled to electromagnetic modes described by Maxwell’s equations, have been justified theoretically \cite{miller_semiclassical_2001, cotton_symmetrical_2013} and applied successfully to problems involving a wide range of light-matter coupling strengths \cite{sukharev_optics_2017, bustamante_dissipative_2021, tancogne-dejean_octopus_2020, yamada_time-dependent_2018, chen_classical_2010, chen_predictive_2019}, including model treatments of molecular polaritons \cite{rosenzweig_analysis_2022}. The applicability of a semiclassical treatment to the strong coupling regime is further motivated by work showing that a formula for the Rabi splitting characteristic of polaritonic absorption spectra can be derived from classical linear dispersion relations without the use of quantum optics \cite{zhu_vacuum_1990}. Moreover, classical transfer matrix methods can accurately predict experimental absorption lineshapes for polaritonic systems \cite{schwennicke_when_2024}. Recent theoretical work has shown that in the thermodynamic limit, absorption spectra predicted by classical transfer matrix methods will match those predicted by quantum optics \cite{yuen-zhou_linear_2024}. The equivalent result in the opposite single-molecule limit, however, remains unproven to the best of our knowledge. In light of the absence of such a proof, it is noteworthy that recent work has suggested some potential limitations of the application of a semiclassical treatment to the strong coupling regime, at least when considering model systems \cite{flick_atoms_2017, ruggenthaler_quantum-electrodynamical_2014, flick_kohnsham_2015, ke_quantum_2024, fiechter_how_2023, simko_twin-polaritons_2025}. Of particular interest here is work suggesting that for a model Hamiltonian, quantum deviations from the classical result for the absorption spectrum are obtained for very small numbers of emitters \cite{zeb_exact_2018}. \par 

The goal of this paper is to compare the classical and quantum mechanical treatments of cavity modes in dynamical first-principles simulations of molecular polaritons and to quantify light-matter entanglement. To enable this comparison, we have developed a hierarchy of real-time propagation methods coupling molecules treated with electronic structure or NEO methods to cavity modes. The ESC regime is studied with conventional electronic RT-TDDFT, and the VSC regime is studied with the RT-NEO approach.  In the semiclassical methods, denoted sc-RT-TDDFT and sc-RT-NEO, the cavity modes are treated classically and are propagated self-consistently with the molecular degrees of freedom through the classical cavity mode coordinate and the expectation value of the molecular dipole moment. To the best of our knowledge, the sc-RT-TDDFT method has not been used previously to study polaritons with conventional electronic structure theory, outside of the NEO framework. In the mean-field-quantum methods, denoted mfq-RT-TDDFT and mfq-RT-NEO, the classical cavity modes are replaced with quantized cavity modes that are propagated self-consistently while coupled to the molecular degrees of freedom through the expectation values of the molecular dipole operator and the coordinate operator of the cavity mode. These mean-field-quantum methods serve as a stepping stone to the full-quantum methods, denoted fq-RT-TDDFT and fq-RT-NEO. 

Similar to the mean-field-quantum methods, the full-quantum methods treat the cavity modes quantum mechanically. However, in contrast to the mean-field-quantum methods, where the molecular and mode subsystems are propagated as coupled but separate systems, the full-quantum methods propagate the two subsystems together as a joint molecule-mode system described by a single density matrix. This approach permits the possibility of quantum entanglement between the molecular and mode subsystems to occur during the dynamics. This entanglement can be quantified at each time step by measuring the von Neumann entropy of either subsystem. We emphasize that quantum entanglement between the molecular and mode subsystems is different from the hybridization that is required to form a hybrid light-matter polaritonic state. Although polariton formation via hybridization occurs in simulations using the semiclassical, mean-field quantum, and full-quantum methods, quantum entanglement can only occur in simulations using the full-quantum method.\par 

The paper is organized as follows. In section 2, we present the theory underlying our hierarchy of real-time methods. Simulation details are provided in section 3. In section 4, we apply our hierarchy to simulate the dynamics of a single H\textsubscript{2} molecule under ESC and a single HCN molecule under VSC. Finally, section 5 presents a more detailed analysis of our findings for H\textsubscript{2} under ESC, focusing on the quantum entanglement.  Conclusions are provided in section 6. \par 

\section{Theory}
\subsection{QED Hamiltonian under the Long-Wavelength Approximation}
We start from the QED Hamiltonian for light-matter interactions written under the long-wavelength approximation:
\begin{equation}
    \hat{H}_{\rm QED} = \hat{H}_{\rm M} + \sum_{k,\lambda}\left[\frac{1}{2}\hat{p}_{k,\lambda}^2 + \frac{1}{2}\omega_{k,\lambda}^2\left( \hat{q}_{k,\lambda} + \frac{\hat{\mu}_{\rm M}\cdot\hat{\xi}_\lambda}{\omega_{k,\lambda}\sqrt{\Omega\epsilon_0}}\right)^2\right]
    \label{qed hamiltonian}
\end{equation}
Here, we define $\hat{H}_{\rm M}$ as the molecular Hamiltonian
\begin{equation}
    \hat{H}_{\rm M} = \sum_i\left(\frac{\hat{\boldsymbol{p}}^2_i}{2m_i} + \hat{V}(\{\hat{\boldsymbol{r}}_i\}, \{\textbf{R}_{\rm C}\})\right)
    \label{molecular hamiltonian}
\end{equation}
where the index $i$ runs over all quantum mechanical particles (i.e., electrons, as well as select quantum nuclei for NEO methods). The operator $\hat{V}(\{\hat{\boldsymbol{r}}_i\}, \{\textbf{R}_{\rm C}\})$ denotes all Coulomb interactions among quantum particles with coordinate operators $\{\hat{\boldsymbol{r}}_i\}$ and classical nuclei at fixed coordinates $\{\textbf{R}_{\rm C}\}$. The summation in Eq. \ref{qed hamiltonian} is over all cavity modes. A given cavity mode is determined by a wave vector $\Vec{k}$ with magnitude $|\Vec{k}| = k = \omega_{k, \lambda}/c$, where $c$ is the speed of light, and polarization vector $\hat{\xi}_\lambda$ satisfying $\Vec{k} \cdot \hat{\xi}_\lambda = 0$. For a cavity mode with wave vector of magnitude $k$ and polarization in the direction of the unit vector $\hat{\xi}_{\lambda}$, $\hat{q}_{k,\lambda}$, $\hat{p}_{k,\lambda}$ and $\omega_{k,\lambda}$ are the coordinate operator, momentum operator, and frequency, respectively. $\Omega$ denotes the effective volume of the cavity, and $\epsilon_0$ is the permittivity of free space. The cavity modes couple linearly to the molecular dipole operator $\hat{\mu}_{\rm M}$. \par
We then expand the square in Eq. \ref{qed hamiltonian} and drop the self-dipole term that is quadratic in $\hat{\mu}_{\rm M}$. This approximation is justified as long as the system remains in the strong coupling regime and does not progress into the ultrastrong coupling  regime\cite{schafer_relevance_2020}, which will be true for all calculations herein. Since the self-dipole term is molecular in character, we do not expect this approximation to significantly impact our analysis of quantum entanglement between the molecule and the cavity mode. The resulting QED Hamiltonian is
\begin{equation}
    \hat{H}_{\rm QED} = \hat{H}_{\rm M} + \hat{H}_{\rm F} + \sum_{k,\lambda}\varepsilon_{k,\lambda}\hat{q}_{k,\lambda}\hat{\mu}_{\rm M}
    \label{bilinear hamiltonian}
\end{equation}
where we have introduced the field Hamiltonian $\hat{H}_{\rm F}$ as the sum of harmonic oscillator Hamiltonians describing the free oscillations of the cavity modes
\begin{equation}
    \hat{H}_{\rm F} = 
    \sum_{k,\lambda}\left[\frac{1}{2}\hat{p}_{k,\lambda}^2 + \frac{1}{2}\omega_{k,\lambda}^2\hat{q}_{k,\lambda}^2\right]
    \label{free field hamiltonian}
\end{equation}
and the light-matter coupling between the molecule and the mode determined by $\{k, \lambda\}$ as
\begin{equation}
    \varepsilon_{k,\lambda} = \frac{\omega_{k,\lambda}}{\sqrt{\Omega\epsilon_0}}
    \label{coupling definition}
\end{equation}
\par
At this point, we must note that different operators in Eq. \ref{bilinear hamiltonian} act on different subspaces of the joint field-molecule Hilbert space $\mathcal{H}_{\rm FM}$, which we will also refer to as the molecule-mode Hilbert space. While $\hat{H}_{\rm M}$ and $\hat{\mu}_{\rm M}$ act on the molecular Hilbert space $\mathcal{H}_{\rm M}$, $\hat{q}_{k,\lambda}$ and $\hat{p}_{k,\lambda}$ act on the mode Hilbert space $\mathcal{H}_{\rm F}$. To emphasize this point and anticipate writing the equations of motion for our various levels of theory, we will rewrite Eq. \ref{bilinear hamiltonian} as
\begin{equation}
    \hat{H}_{\rm QED} = \hat{I}_{\rm F} \otimes \hat{H}_{\rm M} + \hat{H}_{\rm F} \otimes \hat{I}_{\rm M} + \sum_{k,\lambda}\varepsilon_{k,\lambda}\hat{q}_{k,\lambda}\otimes\hat{\mu}_{\rm M}
    \label{bilinear hamiltonian joint hilbert space}
\end{equation}
where we have introduced $\hat{I}_{\rm F(M)}$ as the identity operator in the mode (molecular) Hilbert space $\mathcal{H}_{\rm F (M)}$ and $\otimes$ denotes the Kronecker product between operators in different Hilbert spaces. The effect of multiplying an operator by $\hat{I}_{\rm F(M)}$ with the Kronecker product is to expand it from the mode (molecular) Hilbert space $\mathcal{H}_{\rm F (M)}$ into the joint molecular-mode Hilbert space $\mathcal{H}_{\rm FM}$. Note that we have adopted the ordering of mode operators multiplied by molecular operators.
\subsection{Full-Quantum Equations of Motion}
In the full-quantum methods, fq-RT-TDDFT and fq-RT-NEO, we use $\hat{H}_{\rm QED}$ as given in Eq. \ref{bilinear hamiltonian joint hilbert space} to propagate a von Neumann equation for a joint molecule-mode density matrix $\textbf{P}_{\rm Fe(n)}(t)$ under ESC (VSC). Here the subscripts e and n refer to electrons and nuclei, respectively. Herein, the electronic molecular degrees of freedom are treated quantum mechanically under both ESC and VSC, whereas the nuclear molecular degrees of freedom are only treated quantum mechanically under VSC. In the fq-RT-TDDFT method applied to the ESC regime, we use conventional electronic structure theory and treat all nuclei as classical point charges. A single equation of motion is then propagated for the joint density matrix $\textbf{P}_{\rm Fe}(t)$:
\begin{equation}
    i\hbar\frac{\partial}{\partial t}\textbf{P}_{\rm Fe}(t) = \left[\textbf{I}_{\rm F} \otimes \textbf{F}_{\rm e}(t) + \textbf{H}_{\rm F} \otimes \textbf{I}_{\rm e} + \sum_{k,\lambda}\Tilde{\varepsilon}_{k,\lambda}\Tilde{\textbf{q}}_{k,\lambda}\otimes\left(\boldsymbol{\mu}_{\rm e} - \mu_{\rm e,0}\textbf{I}_{\rm e}\right), \,\textbf{P}_{\rm Fe}(t) \right]
    \label{fq rt tddft eom}
\end{equation}
Here, $\textbf{I}_{\rm F}$ and $\textbf{H}_{\rm F}$ are the matrix representations of the operators $\hat{I}_{\rm F}$ and $\hat{H}_{\rm F}$ defined previously. $\textbf{I}_{\rm e}$ and $\boldsymbol{\mu}_{\rm e}$ are the matrix representations of the identity and dipole operators, respectively, in the electronic Hilbert space. The electronic Kohn-Sham matrix $\textbf{F}_{\rm e}(t)$ is given by 
\begin{equation}
    \textbf{F}_{\rm e}(t) = \textbf{H}^{\rm e}_{\rm core} + \textbf{J}^{\rm ee}[\textbf{P}_{\rm e}(t)] + \textbf{V}^{\rm ee}_{\rm xc}[\textbf{P}_{\rm e}(t)]
    \label{fq rt tddft elec KS}
\end{equation}
Here, $\textbf{H}^{\rm e}_{\rm core}$ is the core Hamiltonian matrix describing electronic kinetic energy and the Coulomb interaction between electrons and classical nuclei; $\textbf{J}^{\rm ee}[\textbf{P}_{\rm e}(t)]$ describes the Coulomb interaction between electrons; and $\textbf{V}^{\rm ee}_{\rm xc}[\textbf{P}_{\rm e}(t)]$ describes the exchange-correlation interaction between electrons. The electronic density matrix $\textbf{P}_{\rm e}(t)$ is defined below. \par
In the summation over cavity modes, we introduce two definitions: $\Tilde{\varepsilon}_{k,\lambda} \equiv \sqrt{2}\varepsilon_{k,\lambda}$ and $\Tilde{\boldsymbol{q}}_{k,\lambda} = \frac{1}{\sqrt{2}}\boldsymbol{q}_{k,\lambda}$, where $\boldsymbol{q}_{k,\lambda}$ is the matrix representation of $\hat{q}_{k,\lambda}$, and the scaling of $\boldsymbol{q}_{k,\lambda}$ manifests in its initial condition. 
The scaling by a factor of $\sqrt{2}$ is appropriate for a restricted Kohn-Sham calculation, where each spatial orbital is doubly occupied and the electronic density matrix only explicitly describes electrons of a single spin. These equations are only applicable to a system composed of two paired electrons, and future work will present the generalization to an arbitrary number of electrons. 
Additionally, we subtract the permanent dipole moment $\mu_{\rm e,0}$ from the dipole moment operator $\boldsymbol{\mu}_{\rm e}$ to ensure that the cavity modes are not excited at time $t = 0$ by the permanent molecular dipole moment. We initialize $\textbf{P}_{\rm Fe}(0)$ as the Kronecker product of a mode density matrix $\textbf{P}_{\rm F}(0)$ and an electronic density matrix $\textbf{P}_{\rm e}(0)$: $\textbf{P}_{\rm Fe}(0) = \textbf{P}_{\rm F}(0) \otimes \textbf{P}_{\rm e}(0)$. The joint electronic-mode density matrix is thus separable into two subsystem density matrices at $t = 0$. \par

In the fq-RT-NEO method applied to the VSC regime, we employ the NEO method to treat the quantized nuclei strongly coupled to the cavity mode(s). In the standard RT-NEO framework without coupling to cavity modes, the total wavefunction $\Psi(\textbf{x}_{\rm e}, \textbf{x}_{\rm n}; t)$ is assumed to be separable into electronic and nuclear parts:
\begin{equation}
    \Psi(\textbf{x}_{\rm e}, \textbf{x}_{\rm n}; t) = \Psi_{\rm e}(\textbf{x}_{\rm e}; t)\Psi_{\rm n}(\textbf{x}_{\rm n}; t)
    \label{joint wavefunction}
\end{equation}
Each of the individual electronic and nuclear wavefunctions is propagated by separate equations of motion. Generalizing the RT-NEO approach to include cavity modes in the fq-RT-NEO method for VSC, we propagate two equations of motion, one for the joint nuclear-mode density matrix $\textbf{P}_{\rm Fn}(t)$ and the other for the electronic density matrix $\textbf{P}_{\rm e}(t)$:
\begin{equation}
    i\hbar\frac{\partial}{\partial t}\textbf{P}_{\rm Fn}(t) = \left[\textbf{I}_{\rm F} \otimes \textbf{F}^{\rm{NEO}}_{\rm n}(t) + \textbf{H}_{{\rm F}} \otimes \textbf{I}_{\rm n} + \sum_{k,\lambda}\varepsilon_{k,\lambda} \textbf{q}_{k,\lambda}\otimes\left(\boldsymbol{\mu}_{\rm n} - \mu_{\rm n,0}\textbf{I}_{\rm n}\right), \,\textbf{P}_{\rm Fn}(t) \right]
    \label{rt neo nuc eom}
\end{equation}
\begin{equation}
        i\hbar\frac{\partial}{\partial t}\textbf{P}_{\rm e}(t) = \left[\textbf{F}^{\rm{NEO}}_{\rm e}(t), \,\textbf{P}_{\rm e}(t) \right]
        \label{rt neo elec eom}
\end{equation}
In Eq. \ref{rt neo nuc eom}, $\textbf{I}_{\rm n}$ and $\boldsymbol{\mu}_{\rm n}$ are the matrix representations of the identity and dipole operators, respectively, in the quantum nuclear Hilbert space. Analogous to the fq-RT-TDDFT equations, we subtract the permanent quantum nuclear dipole moment $\mu_{\rm n,0}$ from the dipole moment operator $\boldsymbol{\mu}_{\rm n}$ to ensure that the cavity modes are not excited at time $t = 0$ by the permanent molecular dipole moment. The time-dependent Kohn-Sham matrices $\textbf{F}^{\rm{NEO}}_{\rm e}(t)$ and $\textbf{F}^{\rm{NEO}}_{\rm n}(t)$ are given by
\begin{equation}
    \textbf{F}^{\rm{NEO}}_{\rm e}(t) = \textbf{H}^{\rm e}_{\rm core} + \textbf{J}^{\rm ee}[\textbf{P}_{\rm e}(t)] + \textbf{V}_{\rm xc}^{\rm ee}[\textbf{P}_{\rm e}(t)] - \textbf{J}^{\rm en}[\textbf{P}_{\rm n}(t)] -  \textbf{V}_{\rm c}^{\rm en}[\textbf{P}_{\rm e}(t),\textbf{P}_{\rm n}(t)] 
    \label{rt neo elec KS}
\end{equation}
\begin{equation}
    \textbf{F}^{\rm{NEO}}_{\rm n}(t) = \textbf{H}^{\rm n}_{\rm core}+ \textbf{J}^{\rm nn}[\textbf{P}_{\rm n}(t)] + \textbf{V}^{\rm nn}_{\rm xc}[\textbf{P}_{\rm n}(t)] - \textbf{J}^{\rm ne}[\textbf{P}_{\rm e}(t)] -  \textbf{V}_{\rm c}^{\rm ne}[\textbf{P}_{\rm n}(t),\textbf{P}_{\rm e}(t)] 
    \label{rt neo nuc KS}
\end{equation}
Here, $\textbf{H}^{\rm n}_{\rm core}$ is the core Hamiltonian describing quantum nuclear kinetic energy and the Coulomb interaction between quantum and classical nuclei; $\textbf{J}^{\rm nn}[\textbf{P}_{\rm n}(t)]$ describes the Coulomb interaction between quantum nuclei; $\textbf{V}^{\rm nn}_{\rm xc}[\textbf{P}_{\rm n}(t)]$ describes the exchange-correlation interaction between quantum nuclei; $\textbf{J}^{\rm ne}[\textbf{P}_{\rm e}(t),\textbf{P}_{\rm n}(t)]$ describes the Coulomb interaction between quantum nuclei and electrons; and $\textbf{V}_{\rm c}^{\rm ne}[\textbf{P}_{\rm e}(t),\textbf{P}_{\rm n}(t)]$ describes the correlation energy between quantum nuclei and electrons. All other variables in Eqs. \ref{rt neo nuc eom} --- \ref{rt neo nuc KS} are defined previously. Note that the quantum nuclei are treated as high spin, with each quantum nucleus occupying its own orbital, and therefore we do not need to introduce any scaled quantities as we did in the ESC equations. These equations are only applicable to the case of one quantized proton, and future work will present the generalization to an arbitrary number of quantum protons. We initialize $\textbf{P}_{\rm Fn}(0)$ as the Kronecker product of a mode density matrix $\textbf{P}_{\rm F}(0)$ and a quantum nuclear density matrix $\textbf{P}_{\rm n}(0)$: $\textbf{P}_{\rm Fn}(0) = \textbf{P}_{\rm F}(0) \otimes \textbf{P}_{\rm n}(0)$. Analogous to $\textbf{P}_{\rm Fe}(0)$ in the fq-RT-TDDFT method, the quantum nuclear-mode density matrix is separable into two subsystem density matrices at $t = 0$. Further discussion of the computation of $\textbf{P}_{\rm F}(0)$ and $\textbf{P}_{\rm n}(0)$ is provided in Simulation Details. \par
At each timestep in the fq-RT-TDDFT (fq-RT-NEO) calculation, we need to use the joint molecule-mode density matrix $\textbf{P}_{\rm Fe}(t)$ ($\textbf{P}_{\rm Fn}(t)$) to obtain the subsystem density matrix describing the molecular degrees of freedom, which are entangled with the cavity degrees of freedom. The corresponding subsystem density matrix is $\textbf{P}_{\rm e}(t)$ in the fq-RT-TDDFT calculation and $\textbf{P}_{\rm n}(t)$ in the fq-RT-NEO calculation. These matrices are needed to compute the electronic (quantum nuclear) Kohn-Sham matrix $\textbf{F}_{\rm e}(t)$ ($\textbf{F}^{\rm{NEO}}_{\rm n}(t)$).
At a given timestep, we can compute the molecular subsystem density matrix, as well as the cavity mode density matrix $\textbf{P}_{\rm F}(t)$, by taking the appropriate partial trace of the joint density matrix $\textbf{P}_{\rm Fe(n)}(t)$:
\begin{equation}
    \textbf{P}_{\rm e(n)}(t) = \textrm{Tr}_{\rm F}[\textbf{P}_{\rm Fe(n)}(t)]
    \label{field partial trace}
\end{equation}
\begin{equation}
    \textbf{P}_{\rm F}(t) = \textrm{Tr}_{\rm e(n)}[\textbf{P}_{\rm Fe(n)}(t)]
    \label{matter partial trace}
\end{equation}
where $\textrm{Tr}_{\rm F}$ denotes a partial trace over the mode degrees of freedom and $\textrm{Tr}_{\rm e(n)}$ denotes a partial trace over electronic (quantum nuclear) degrees of freedom. These subsystem density matrices can also be used to compute observables of the individual subsystems at each time step. \par

In this formulation, we have made the approximation that under VSC, only the quantum nuclear degrees of freedom are directly coupled to the cavity mode(s). The interaction between the cavity mode(s) and the electrons occurs indirectly through the dependence of the quantum nuclear Kohn-Sham matrix $\textbf{F}_{\rm n}(t)$ on the electronic density matrix $\textbf{P}_{\rm e}(t)$, as given by Eq. \ref{rt neo elec KS}. This approximation is expected to be justified on the grounds that the large energy difference between quantum nuclear and electronic transitions makes the effect of coupling to $\boldsymbol{\mu}_{\rm e}$ negligible. Moreover, this formulation applies to experimental measurements on the electronic ground state, where infrared radiation probes vibrational polaritons. \par

We note that the fq-RT-NEO approach with quantized nuclei can also be applied to the ESC regime when the cavity mode frequency corresponds to an electronic excitation, as in a previous application of the semiclassical RT-NEO approach to photoexcited proton transfer in a cavity \cite{li_semiclassical_2022}. The equations of motion for this case are obtained by switching the subscripts e and n in Eqs. \ref{rt neo nuc eom} --- \ref{rt neo elec eom}. The electrons and quantized mode are allowed to entangle, while the quantum nuclei are propagated with a separate von Neumann equation. The quantum nuclei interact indirectly with the mode through the dependence of the electronic Kohn-Sham matrix $\textbf{F}_{\rm e}(t)$ on the quantum nuclear density matrix $\textbf{P}_{\rm n}(t)$. \par

\subsection{von Neumann Entropy}
In the full-quantum methods, we initialize the joint molecule-mode density matrix as a separable product of two pure state density matrices, one for the mode subsystem and the other for the molecular subsystem. As we propagate the evolution of the joint density matrix according to Eq. \ref{fq rt tddft eom} under ESC or Eq. \ref{rt neo nuc eom} under VSC, it generally becomes impossible to write the joint density matrix as a product of two separable pure state subsystem density matrices at any given time $t > 0$. Each subsystem will evolve from a pure state to a mixed state that must be described by a statistical ensemble. In this case, the molecular and mode subsystems have become entangled with one another. As the subsystems become more entangled, the quantum states of the subsystems move further away from being in pure states. Quantifying the degree of entanglement of the subsystems is therefore equivalent to quantifying how close the subsystems are to being in pure states. \par
To quantify how close the subsystem density matrices are to being in pure states at a given time $t$, we use the von Neumann entropy $S(t)$. For any density matrix $\textbf{P}(t)$, the von Neumann entropy is defined as
\begin{equation}
    S(t) = -\textrm{Tr}(\textbf{P}(t)\ln\textbf{P}(t))
    \label{von neumann}
\end{equation}
As discussed below, this quantity provides a measure of how close the density matrix $\textbf{P}(t)$ is to being a pure state, or equivalently, the degree to which $\textbf{P}(t)$ is in a mixed state. In the case that $\textbf{P}(t)$ describes either the molecule or mode subsystem density matrices, $S(t)$ quantifies how entangled the molecule and mode subsystems are.\par
We will now enumerate some properties of the von Neumann entropy, which will be useful for interpreting our results. Further references and more detailed proofs of these statements are provided in the Supporting Information, and the interested reader may also consult Ref. \citenum{nielsen_quantum_2009}.
\begin{enumerate}
    \item $S(t)$ is always non-negative and has a minimum possible value of 0 corresponding to a pure state. It is maximized for a random statistical ensemble; if the ensemble has $N$ replicas, then the maximum possible value of $S(t)$ is $S_{\rm max} = \ln(N)$. 
    \item $S(t)$ is invariant under a unitary transformation. Taken together, statements 1 and 2 tell us that a system initialized in a pure state has zero von Neumann entropy at $t = 0$ and will continue to have zero von Neumann entropy (remain in a pure state) under unitary time evolution described by a von Neumann equation. In the full-quantum methods, this statement applies to the joint density matrices $\textbf{P}_{\rm Fe}(t)$ and $\textbf{P}_{\rm Fn}(t)$, as well as the electronic density matrix $\textbf{P}_{\rm e}(t)$ in the fq-RT-NEO method. These density matrices evolve according to the von Neumann Eqs. \ref{fq rt tddft eom}, \ref{rt neo nuc eom}, and \ref{rt neo elec eom}, respectively. Since they are initialized as pure state density matrices and their time evolution is governed by a von Neumann equation, these density matrices will remain in pure states at all times. In contrast, the molecular subsystem density matrices computed by taking a partial trace of the joint molecule-mode density matrix, as given in Eqs. \ref{field partial trace} and \ref{matter partial trace}, do not evolve according to a von Neumann equation. They start as pure states at $t = 0$ but evolve into mixed states over time.
    \item Consider a system AB composed of subsystems A and B and described by a density operator $\hat{\rho}_{\rm AB}(t)$. We can compute the von Neumann entropy for the total system AB, $S_{\rm AB}(t)$, according to Eq. \ref{von neumann}: $S_{\rm AB}(t) = -\textrm{Tr}(\hat{\rho}_{\rm AB}(t)\ln\hat{\rho}_{\rm AB}(t))$. We can also compute the von Neumann entropies $S_{\rm A}(t)$ and $S_{\rm B}(t)$ for each of the two subsystems A and B. If subsystem A (B) is described by a density operator $\hat{\rho}_{\rm A(B)}(t)$, then $S_{\rm A(B)}(t) = -\textrm{Tr}(\hat{\rho}_{\rm A(B)}(t)\ln\hat{\rho}_{\rm A(B)}(t))$. The three von Neumann entropies $S_{\rm AB}(t)$, $S_{\rm A}(t)$, and $S_{\rm B}(t)$ are related by the inequality 
    \begin{equation}
        S_{\rm AB}(t) \leq S_{\rm A}(t) + S_{\rm B}(t)
        \label{subadditivity}
    \end{equation}
    This property is known as \textit{subadditivity}. Equality holds if $\hat{\rho}_{\rm AB}(t)$ is separable: $\hat{\rho}_{\rm AB}(t) = \hat{\rho}_{\rm A}(t) \otimes \hat{\rho}_{\rm B}(t)$. Physically, it means that the two subsystems A and B are not entangled with each other and are independent. \par
    In the context of our full-quantum methods, the joint molecule-mode density matrix is analogous to the density operator $\hat{\rho}_{\rm AB}(t)$, and the individual mode and molecular subsystem density matrices are analogous to the density operators $\hat{\rho}_{\rm A}(t)$ and $\hat{\rho}_{\rm B}(t)$. We established from statement 2 that the joint molecule-mode system is always in a pure quantum state, so its von Neumann entropy $S_{\rm MF}(t)$ will always be zero. The entropies of the molecule and mode subsystems $S_{\rm M}(t)$ and $S_{\rm F}(t)$, however, will in general be positive numbers. These nonzero subsystem entropies will arise as the two subsystems entangle, with each entering a mixed quantum state. \par 
    \item For the system AB consisting of two entangled subsystems A and B, we have concluded that we can determine the degree of entanglement of the two subsystems by measuring the von Neumann entropy of one of them. It does not matter which subsystem's von Neumann entropy is measured because the subsystem entropies $S_{\rm A}(t)$ and $S_{\rm B}(t)$ defined in statement 3 are equal at all times $t$:
    \begin{equation}
        S_{\rm A}(t) = S_{\rm B}(t)
        \label{subspace entropy equality}
    \end{equation}
    This relation holds true \textit{irrespective of whether or not the dimensions of the Hilbert spaces for the two subsystems are the same}, an important requirement for our work because the molecular and mode subsystems are described with finite-dimensional bases (i.e., atomic orbital basis sets for the molecular subsystem and a set of harmonic oscillator number basis functions for the mode), which are in general not the same size. This statement tells us that it does not matter which subsystem von Neumann entropy we measure to determine the degree of molecular-mode entanglement. In our code, we measure the von Neumann entropy of the molecular subsystem, but we have confirmed that numerically equivalent results are obtained from the mode subsystem.
    \end{enumerate}
\subsection{Mean-Field-Quantum Equations of Motion}
In the mean-field-quantum methods, mfq-RT-TDDFT and mfq-RT-NEO, we assume that the joint molecule-mode density matrix $\textbf{P}_{\rm Fe(n)}(t)$ is separable into mode and molecular pure state density matrices at all times $t \geq 0$:
\begin{equation}
    \textbf{P}_{\rm Fe(n)}(t) = \textbf{P}_{\rm F}(t) \otimes \textbf{P}_{\rm e(n)}(t)
    \label{separability ansatz}
\end{equation}
This assumption is equivalent to assuming that there is zero molecule-mode entanglement. \par

We can plug this ansatz into Eq. \ref{fq rt tddft eom} for the fq-RT-TDDFT method and Eq. \ref{rt neo nuc eom} for the fq-RT-NEO method. By repeatedly applying the mixed-product property of the Kronecker product, we arrive at the equations of motion for the mfq-RT-TDDFT and mfq-RT-NEO methods. The ansatz in Eq. \ref{separability ansatz} has the effect of separating Eq. \ref{fq rt tddft eom} and Eq. \ref{rt neo nuc eom} each into two coupled equations of motion, one for the mode and another for the molecular degrees of freedom strongly coupled to it. Both the molecular and mode subsystems are still treated quantum mechanically and are coupled to each other, but each subsystem now evolves under its own equation of motion. \par
For the mfq-RT-TDDFT method applied to the ESC regime, the resulting equations of motion for the mode density matrix $\textbf{P}_{\rm F}(t)$ and electronic density matrix $\textbf{P}_{\rm e}(t)$ are
\begin{equation}
    i\hbar \frac{\partial}{\partial t}\textbf{P}_{\rm F}(t) = \left[\textbf{H}_{\rm F} + \sum_{k,\lambda}\varepsilon_{k,\lambda}\textbf{q}_{k,\lambda}\langle\boldsymbol{\mu}_{\rm e}\rangle_t, \:\textbf{P}_{\rm F}(t)\right] 
    \label{mf rt tddft mode eom}
\end{equation}
\begin{equation}
    i\hbar \frac{\partial}{\partial t}\textbf{P}_{\rm e}(t) = \left[\textbf{F}_{\rm e}(t) + \sum_{k,\lambda}\varepsilon_{k,\lambda}\langle \textbf{q}_{k,\lambda} \rangle_t\boldsymbol{\mu}_{\rm e}, \:\textbf{P}_{\rm e}(t)\right] 
    \label{mf rt tddft electronic eom}
\end{equation}
Here, $\langle\textbf{q}_{k,\lambda}\rangle_t \equiv \textrm{Tr}(\textbf{P}_{\rm F}(t)\textbf{q}_{k,\lambda})$ denotes the expectation value of $\hat{q}_{k,\lambda}$ at time $t$, and $\langle\boldsymbol{\mu}_{\rm e}\rangle_t \equiv 2\textrm{Tr}(\textbf{P}_{\rm e}(t)\boldsymbol{\mu}_{\rm e}) - 2\textrm{Tr}(\textbf{P}_{\rm e}(0)\boldsymbol{\mu}_{\rm e})$ denotes the time-dependent electronic dipole moment at time $t$ minus the permanent dipole moment. The prefactor of 2 in front of the trace allows us to account for electrons of both spins, eliminating the need to use the scaled quantities in Eq. \ref{fq rt tddft eom}.  Thus, the mean-field quantum equations are valid for any number of paired electrons. \par
For the mfq-RT-NEO method applied to the VSC regime, the resulting equations of motion for the mode density matrix $\textbf{P}_{\rm F}(t)$, quantum nuclear density matrix $\textbf{P}_{\rm n}(t)$, and electronic density matrix $\textbf{P}_{\rm e}(t)$ are
\begin{equation}
    i\hbar \frac{\partial}{\partial t}\textbf{P}_{\rm F}(t) = \left[\textbf{H}_{\rm F} + \sum_{k,\lambda}\varepsilon_{k,\lambda}\textbf{q}_{k,\lambda}\langle\boldsymbol{\mu}_{\rm n}\rangle_t, \:\textbf{P}_{\rm F}(t)\right]
    \label{mf rt neo field eom}
\end{equation}
\begin{equation}
    i\hbar \frac{\partial}{\partial t}\textbf{P}_{\rm n}(t) = \left[\textbf{F}_{\rm n}(t) + \sum_{k,\lambda}\varepsilon_{k,\lambda}\langle\textbf{q}_{k,\lambda}\rangle_t\boldsymbol{\mu}_{\rm n}, \:\textbf{P}_{\rm n}(t)\right] 
    \label{mf rt neo nuc eom}
\end{equation}
\begin{equation}
    i\hbar\frac{\partial}{\partial t}\textbf{P}_{\rm e}(t) = \left[\textbf{F}_{\rm e}(t), \,\textbf{P}_{\rm e}(t) \right]
    \label{mf rt neo elec eom}
\end{equation}
Here, $\langle\boldsymbol{\mu}_{\rm n}\rangle_t \equiv \textrm{Tr}(\textbf{P}_{\rm n}(t)\boldsymbol{\mu}_{\rm n}) - \textrm{Tr}(\textbf{P}_{\rm n}(0)\boldsymbol{\mu}_{\rm n})$ denotes the time-dependent quantum nuclear dipole moment at time $t$ minus the permanent dipole moment. \par

Eqs. \ref{mf rt neo field eom} --- \ref{mf rt neo elec eom} assume that the cavity modes are only coupled directly to the quantum nuclear dipole moment. As discussed in Section 2.2, the coupling of the cavity mode to the electronic dipole moment under VSC is expected to be negligible due to the large energy difference between vibrational and electronic transitions. In Section S2 of the Supporting Information, we evaluate the effects of this assumption by comparing the results obtained with this mfq-RT-NEO method to the results obtained with what is denoted the total-coupling mfq-RT-NEO method, where the cavity modes are coupled directly to both the electronic and quantum nuclear dipole moments. This comparison reveals that coupling to both electronic and nuclear dipole moments in the VSC regime may cause unphysical behavior of the polaritonic spectrum, most likely due to a poor description of higher-energy electronic transitions by TDDFT. In general, direct coupling of the cavity modes to only the nuclear dipole moment is appropriate for describing VSC experiments on the electronic ground state, whereas direct coupling of the cavity modes to both the electronic and nuclear dipole moments may be appropriate for VSC experiments on electronic excited states. Our findings suggest that more accurate vibrational polaritonic spectra on the electronic ground state may be obtained by assuming that the cavity modes only couple directly to the quantum nuclear dipole moment and justifies our use of this assumption in our other VSC methods. 

\subsection{Semiclassical Equations of Motion}
The equations of motion for the semiclassical methods, sc-RT-TDDFT and sc-RT-NEO, are obtained by replacing the operators $\hat{q}_{k,\lambda}(t)$ and $\hat{p}_{k,\lambda}(t)$ in the von Neumann equations given by Eqs. \ref{mf rt tddft mode eom} and \ref{mf rt neo field eom} with their corresponding classical coordinates $q_{k,\lambda}(t)$ and $p_{k,\lambda}(t)$. In the semiclassical methods, all molecular degrees of freedom are treated quantum mechanically, while the cavity mode(s) are treated as classical harmonic oscillators. \par
The equations of motion for the sc-RT-TDDFT method under ESC are
\begin{equation}
    \Dot{q}_{k,\lambda}(t) = p_{k,\lambda}(t)
    \label{sc rt tddft mode position eom}
\end{equation}
\begin{equation}
    \Dot{p}_{k,\lambda}(t) = -\omega_{k,\lambda}^2 q_{k,\lambda}(t) - \varepsilon_{k,\lambda}\langle\boldsymbol{\mu}_{\rm e}\rangle_t - \gamma_{\rm c}p_{k,\lambda}(t)
    \label{sc rt tddft mode momentum eom}
\end{equation}
\begin{equation}
    i\hbar \frac{\partial}{\partial t}\textbf{P}_{\rm e}(t) = \left[\textbf{F}_{\rm e}(t) + \sum_{k,\lambda}\varepsilon_{k,\lambda} q_{k,\lambda}(t) \boldsymbol{\mu}_{\rm e}, \:\textbf{P}_{\rm e}(t)\right] 
    \label{sc rt tddft electronic eom}
\end{equation}
In Eq. \ref{sc rt tddft mode momentum eom}, we have phenomenologically introduced a friction term with damping constant $\gamma_{\rm c}$ that generates loss of field intensity in the cavity (the inverse of the decoherence timescale of the field). \par
The equations of motion for the sc-RT-NEO method under VSC are
\begin{equation}
    \Dot{q}_{k,\lambda}(t) = p_{k,\lambda}(t)
    \label{sc rt neo mode position eom}
\end{equation}
\begin{equation}
    \Dot{p}_{k,\lambda}(t) = -\omega_{k,\lambda}^2 q_{k,\lambda}(t) - \varepsilon_{k,\lambda}\langle\boldsymbol{\mu}_{\rm n}\rangle_t - \gamma_{\rm c}p_{k,\lambda}(t)
    \label{sc rt neo mode momentum eom}
\end{equation}
\begin{equation}
    i\hbar \frac{\partial}{\partial t}\textbf{P}_{\rm n}(t) = \left[\textbf{F}_{\rm n}(t) + \sum_{k,\lambda}\varepsilon_{k,\lambda} q_{k,\lambda}(t) \boldsymbol{\mu}_{\rm n}, \:\textbf{P}_{\rm n}(t)\right] 
    \label{sc rt neo nuclear eom}
\end{equation}
\begin{equation}
    i\hbar \frac{\partial}{\partial t}\textbf{P}_{\rm e}(t) = \left[\textbf{F}_{\rm e}(t), \:\textbf{P}_{\rm e}(t)\right] 
    \label{sc rt neo electronic eom}
\end{equation}
The sc-RT-TDDFT and sc-RT-NEO methods in this work are variations of the semiclassical approach implemented in Ref. \citenum{li_semiclassical_2022}. Here, we compare their results to those of the newly implemented mean-field-quantum and full-quantum methods.
\subsection{Electronic Born-Oppenheimer Approximation}
In all of the RT-NEO methods under VSC, we propagate the electronic degrees of freedom simultaneously with the quantum nuclear and field degrees of freedom. Accurate simulation of the fast electron dynamics requires a relatively small time step, which increases the computational cost of the calculation. An alternative approach, which we refer to as the electronic Born-Oppenheimer-RT-NEO (BO-RT-NEO) method \cite{li_electronic_2023}, circumvents this problem by quenching the electronic density to the ground state at each time step. The quantum nuclear degrees of freedom are propagated on an instantaneous potential energy surface determined by the electronic ground state density for the nonequilibrium quantum nuclear density and the positions of the fixed classical nuclei. 
This quenching eliminates the need for a small time step to explicitly propagate the electron dynamics, instead allowing for an approximately order-of-magnitude larger time step and a significant reduction in computational cost. We can apply this electronic Born-Oppenheimer approximation to any of the treatments in the VSC regime given above by replacing the electronic equation of motion with 
\begin{equation}
    \textbf{P}_{\rm e}(t) = \textrm{SCF}\left[\Re(\textbf{P}_{\rm n}(t)), \;\textbf{R}_{\rm c}\right]
    \label{bo scf equation}
\end{equation}
where $\Re(\textbf{P}_{\rm n}(t))$ is the real part of the quantum nuclear density matrix at time $t$, and $\textbf{R}_{\rm c}$ is the vector of coordinates of the fixed classical nuclei. Note that the contribution of the cavity mode appears only indirectly in this electronic equation through the dependence of the nuclear density matrix at time $t$ on the field. The sc-BO-RT-NEO method was previously implemented for polaritonic systems \cite{li_electronic_2023}. Here, we extend the electronic Born-Oppenheimer approximation further to implement both the mfq-BO-RT-NEO and fq-BO-RT-NEO methods.
\section{Simulation Details}
All of the methods listed above have been implemented in a developer version of QChem \cite{epifanovsky_software_2021}. With the exception of the electronic degrees of freedom in the electronic Born-Oppenheimer methods and the quantum modes in the mean-field methods, all quantum mechanical degrees of freedom are propagated with a modified-midpoint unitary transform (MMUT) time-propagation scheme \cite{li_time-dependent_2005} with an additional predictor-corrector scheme \cite{de_santis_pyberthart_2020} used to control numerical error during the real-time dynamics. The classical nuclei are fixed at the specified geometries in all calculations. In the semiclassical methods, the classical modes are propagated with the velocity Verlet algorithm, while in the mean-field-quantum methods, they are propagated with the time-evolution operator $\exp(-i\textbf{H}t/\hbar)$, where $\textbf{H} = \textbf{H}_{\rm F} + \sum_{k,\lambda}\varepsilon_{k,\lambda}\textbf{q}_{k,\lambda}\langle\boldsymbol{\mu}_{\rm e/n}\rangle_t$ as given in Eqs. \ref{mf rt tddft mode eom} and  \ref{mf rt neo field eom}. \par

For the ESC calculations on H\textsubscript{2}, we used the 6-31G \cite{hehre_selfconsistent_1972} electronic basis set and the B3LYP \cite{lee_development_1988, becke_new_1998, becke_density-functional_1988} electronic exchange-correlation functional. For the VSC calculations on HCN, all electrons and the hydrogen nucleus were treated quantum mechanically. We used the cc-pVDZ \cite{dunning_gaussian_1989} electronic basis set and the even-tempered 8s8p8d \cite{yang_development_2017} protonic basis set with exponents ranging from $2\sqrt{2}$ to 32, as well as the B3LYP \cite{lee_development_1988, becke_new_1998, becke_density-functional_1988} electronic exchange-correlation functional and the epc17-2 \cite{yang_development_2017, brorsen_multicomponent_2017} electron-proton correlation functional.
For the mean-field-quantum and full-quantum calculations using quantized modes, we used four harmonic oscillator number basis functions $\ket{i}, i = 0, 1,2,3$ to describe the mode. This number of basis functions was found to be sufficient to generate a converged coherent state for the initial field amplitudes used. 

For the ESC calculations on H\textsubscript{2}, the initial electronic density matrix $\textbf{P}_{\rm e}(0)$ is taken to be the SCF ground state from a conventional electronic DFT calculation with a bond length of 0.74 a.u. For the VSC calculations on HCN, the initial electronic and nuclear density matrices $\textbf{P}_{\rm e}(0)$ and $\textbf{P}_{\rm n}(0)$ are taken to be the SCF ground state from a NEO-DFT calculation with a C-N bond length of 1.16 a.u. and the proton basis function center located 1.07 a.u. from the carbon nucleus. 
For the full-quantum and mean-field-quantum methods, the initial mode density matrix $\textbf{P}_{\rm F}(0)$ is taken to be the density matrix describing the harmonic oscillator ground state $\ket{0}$.
This choice sets the mode position and momentum expectation values to zero. Note that the overlap of the initial state with the exact ground eigenstate obtained by diagonalizing the full QED Hamiltonian at $t$ = 0 is \textgreater 0.99. For the semiclassical methods, the initial position and momentum of the mode is set to zero: $q_{k,\lambda}(0) = p_{k,\lambda}(0) = 0$. \par

At $t = 0$ in all calculations, we apply a delta pulse that excites the mode degrees of freedom. In the semiclassical methods, the delta pulse is realized by setting $q_{k,\lambda} = E_0$, where $E_0$ is the amplitude of the pulse. $E_0 = 0.01$ a.u. for ESC calculations and 0.3 a.u. for VSC calculations to increase the strength of the protonic dipole moment signal. In the mean-field-quantum and full-quantum methods, the delta pulse is realized by setting $\textbf{P}_{\rm F}(0)$ to describe the coherent state $\ket{\rm C}$ created by displacing the harmonic oscillator ground state by $E_0$: $\ket{\rm C} = \exp(-\omega_{k,\lambda}E_0^2/4)\exp\left(\sqrt{\omega_{k,\lambda}/2} E_0\hat{a}^{\dagger}_{k,\lambda}\right)\ket{0}$, where $\hat{a}^{\dagger}_{k,\lambda}$ is the creation operator that adds a quantum of energy to the mode characterized by $\{k,\lambda\}$. This translation in coordinate space leaves the initial expectation value of the mode momentum unchanged: $\braket{{\rm C}|\hat{p}_{k,\lambda}|\rm C} = \braket{0|\hat{p}_{k,\lambda}|0} = 0$. The selection of a coherent state, which would oscillate like a classical harmonic oscillator when uncoupled from the molecule, is intended to facilitate the most even-handed comparison between methods using classical and quantized modes. We recognize that choosing a different initial condition, such as a Fock state, may lead to a different degree of light-matter entanglement. This possibility will be considered in future work. Finally, the joint density matrices $\textbf{P}_{\rm Fe(n)}(0)$ in the full-quantum calculations can then be initialized as the product of two separable subsystem density matrices: $\textbf{P}_{\rm Fe(n)}(0) = \textbf{P}_{\rm F}(0) \otimes \textbf{P}_{\rm e(n)}(0)$.  \par

All ESC semiclassical and full-quantum calculations used the timestep $\Delta{t} = 0.1$ a.u. and light-matter coupling $0.004$ a.u. This timestep was reduced to $\Delta{t} = 0.01$ a.u. for the ESC mean-field-quantum calculation to reduce numerical error. For the VSC calculations on HCN, the timestep was $\Delta{t} = 0.1$ a.u. and the light-matter coupling was $8 \times 10^{-4}$ a.u. All Fourier transforms $\textrm{P}(\omega)$ of real-time data $f(t)$ were computed using the Pad\'{e} approximation $\textrm{P}(\omega) = |\mathcal{F}\left[f(t)e^{-\gamma t}\right]|$, with a small damping of $\gamma = 10^{-5}$ a.u., giving a linewidth of $1.7 \times10^{-3}$ eV = 13.8 cm\textsuperscript{-1} to all peaks. For the linear response spectra shown in Figs. \ref{esc sc mf}b and \ref{fig:fq_vsc}b below, we used a slightly smaller damping of $\gamma = 10^{-6}$ a.u., giving a linewidth of $1.7 \times10^{-4}$ eV = 1.38 cm\textsuperscript{-1} to all peaks in those spectra. This choice was made in order to obtain roughly equal intensities between polariton peaks and does not alter the peak locations. A comparison of the effect of different damping factors for the data in Fig. \ref{esc sc mf}b is shown in Figure S4. \par

\section{Results}

\subsection{Strong coupling with mean-field light-matter interactions}

\begin{figure}[H]
\centering
\includegraphics[scale=0.5]{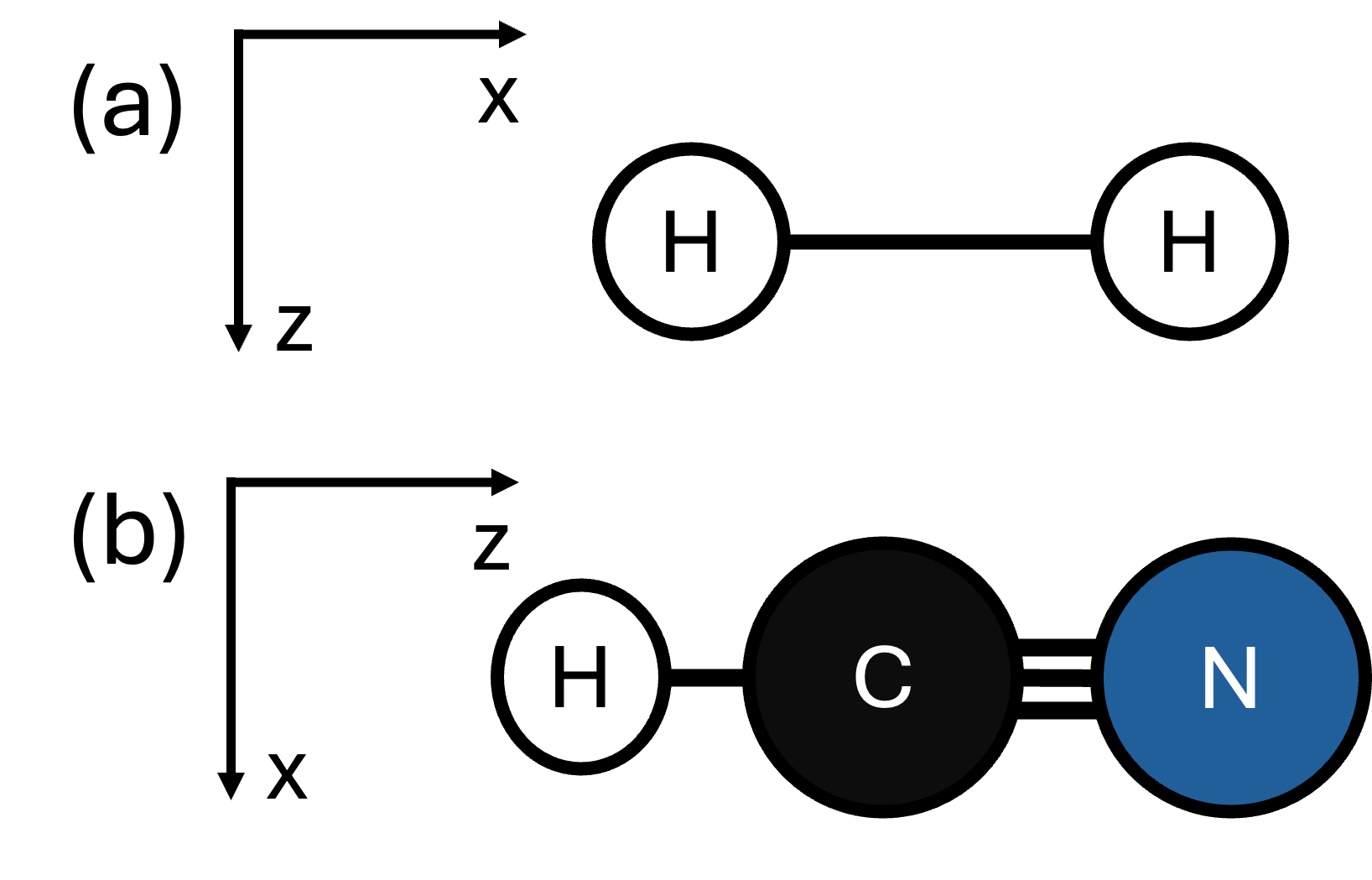}
    \caption{(a) Geometry of the H\textsubscript{2} molecule used in the RT-TDDFT ESC calculations. (b) Geometry of the HCN molecule used in the RT-NEO VSC calculations.}
    \label{MoleculeSetup}
\end{figure}

We first consider a single H\textsubscript{2} molecule oriented as shown in Fig. \ref{MoleculeSetup}a. The classical nuclei are fixed at the specified geometry throughout the dynamics. Since this molecule contains only two electrons of opposite spins, and the electronic density matrix $\textbf{P}_{\rm e}(t)$ is evolved under restricted RT-TDDFT, only a one-electron density matrix is explicitly propagated. To model strong light-matter interactions, the molecule is coupled to an $x$-polarized lossless cavity mode with a coupling strength $\varepsilon = 4 \times 10^{-3}$  a.u. The frequency of the cavity mode is set to $\omega_{\rm c}$ = 14.750 eV, in resonance with the strongest electronic transition of H\textsubscript{2} at this level of theory.  At $t = 0$, the cavity mode is perturbed by a delta pulse, thus driving the excited-state dynamics of the coupled light-matter system. 

\begin{figure}[H]
\centering
\includegraphics[scale=0.4]{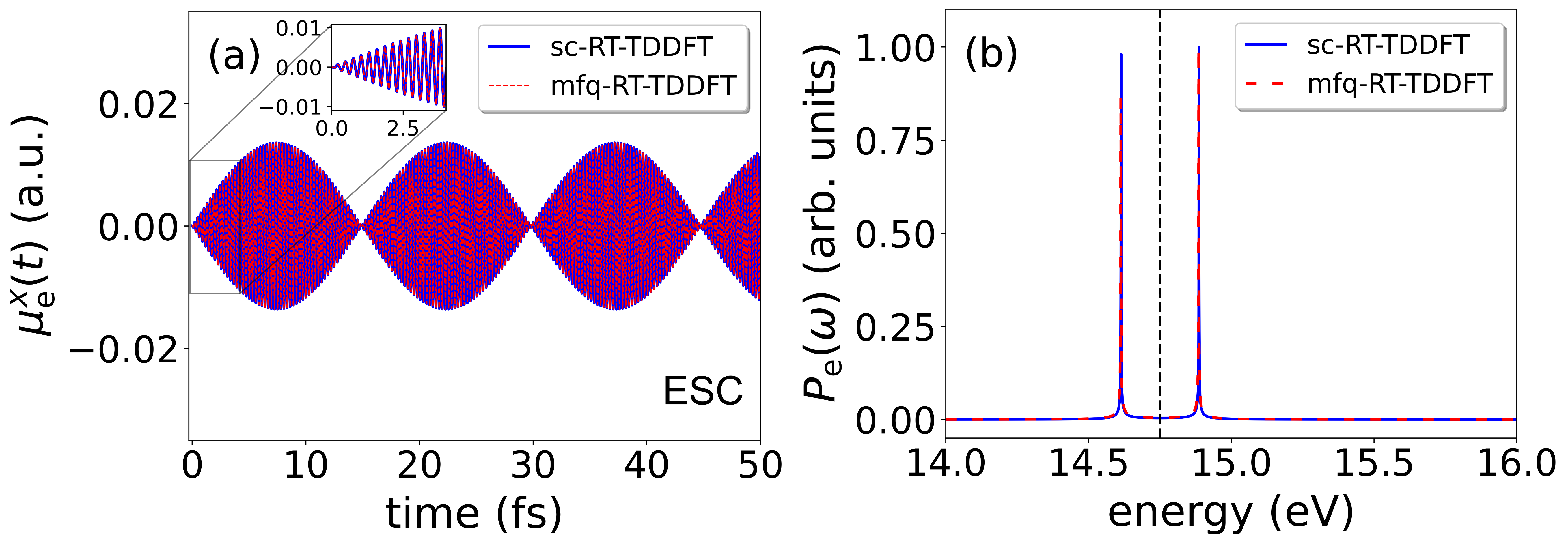}
\caption{(a) sc-RT-TDDFT and mfq-RT-TDDFT dynamics of the $x$-component of the H\textsubscript{2} electronic dipole moment $\mu_{\rm e}^x(t)$, coupled to an $x$-polarized cavity mode with frequency $\omega_{\rm c}$ = 14.750 eV perturbed by a delta pulse at $t = 0$ with coupling strength 4 $\times$ 10\textsuperscript{-3} a.u. (b) Power spectra $P_{\rm e}(\omega)$ corresponding to the dipole signals in panel a, along with $\omega_{\rm c}$ (vertical black dashed line). As expected, the results from the sc-RT-TDDFT and mfq-RT-RDDFT methods are indistinguishable.  }
\label{esc sc mf}
\end{figure}

Fig. \ref{esc sc mf}a shows the evolution of the time-dependent electronic dipole moment in the $x$-direction, $\mu_{\rm e}^x(t) \equiv \langle\hat{\mu}_{\rm e}^x\rangle_t - \langle\hat{\mu}_{\rm e}^x\rangle_0$, evolved with both the sc-RT-TDDFT and mfq-RT-TDDFT methods. Both methods treat the light-matter coupling on the mean-field level, with the former (latter) method modeling the cavity photon mode classically (quantum mechanically).
As expected, the two methods predict virtually identical $\mu_{\rm e}^x(t)$ dynamics. In addition to the fast, sub-femtosecond oscillations due to the electronic transition, Rabi oscillations also occur on a period of around 15 fs. According to Fig. \ref{esc sc mf}b, the power spectra of the time-domain signals reveal a pair of lower polariton (LP) and upper polariton (UP) peaks resulting from the resonance coupling between the cavity mode and the H\textsubscript{2} electronic transition. The polariton peaks are centered at the cavity mode energy of 14.75 eV (black dashed line). The Rabi splitting of the two polariton peaks is $\Omega_{\rm R} = 0.27$ eV. We note that the same Rabi splitting can also be obtained from the Fourier transform of $q(t)$, as shown in Figure S5.\par

\begin{figure}[H]
\centering
\includegraphics[scale=0.4]{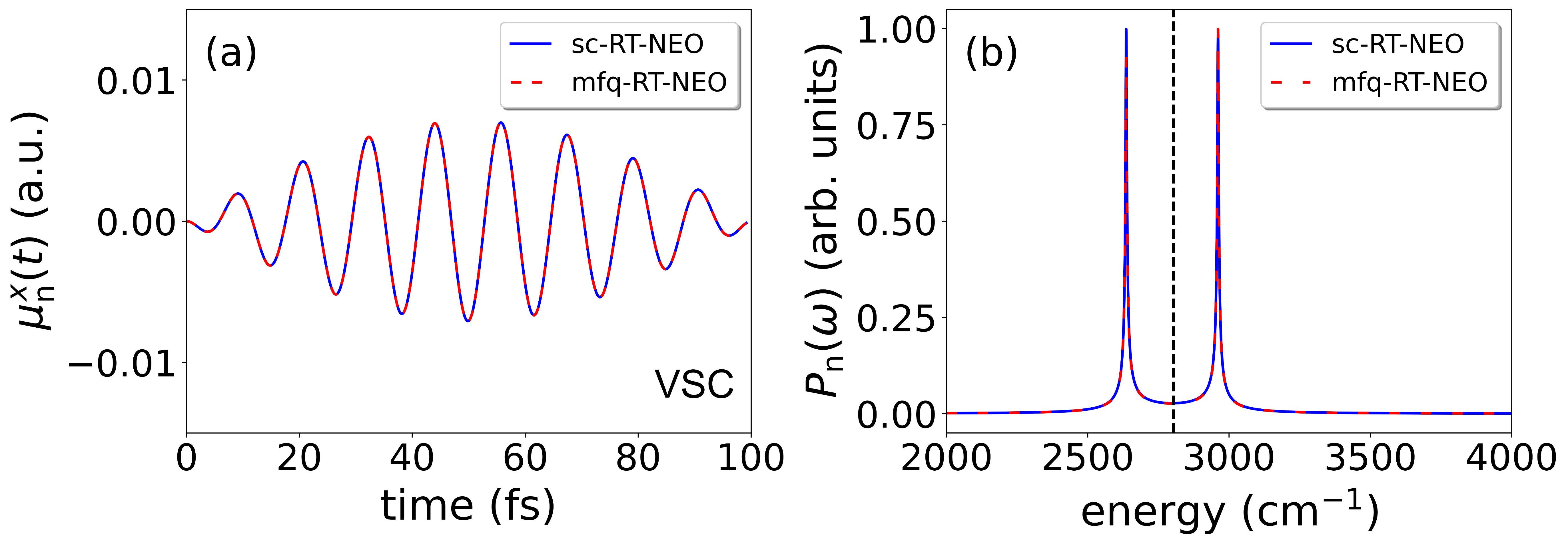}
\caption{(a) sc-RT-NEO and mfq-RT-NEO dynamics of the $x$-component of the HCN nuclear dipole moment $\mu_{\rm n}^x(t)$, coupled to an $x$-polarized cavity mode with frequency $\omega_{\rm c}$ = 2803 cm\textsuperscript{-1} perturbed by a delta pulse at $t = 0$ with coupling strength 8 $\times$ 10\textsuperscript{-4} a.u. (b) Power spectra $P_{\rm n}(\omega)$ corresponding to the dipole signals in panel a, along with $\omega_{\rm c}$ (vertical black dashed line). As expected, the results from the sc-RT-NEO and mfq-RT-NEO methods are indistinguishable.}
\label{vsc sc mf pc}
\end{figure}

After studying the ESC regime, we now consider VSC formed between the protonic motion of a single HCN molecule (see Fig. \ref{MoleculeSetup}b) and an infrared (IR) cavity mode. All electrons and the proton of HCN are treated quantum mechanically with the RT-NEO approach, and the molecule is coupled to an $x$-polarized cavity mode. The frequency of the mode is tuned to $\omega_{\rm c}$ = 2803 cm\textsuperscript{-1}, in resonance with the $x$-direction bending motion of the quantum proton. Fig. \ref{vsc sc mf pc}a shows the evolution of the time-dependent nuclear dipole moment in the $x$-direction, $\mu_{\rm n}^x(t) \equiv \langle\hat{\mu}_{\rm n}^x\rangle_t - \langle\hat{\mu}_{\rm n}^x\rangle_0$, when the cavity mode is perturbed at $t = 0$ by a delta pulse. Both the sc-RT-NEO and mfq-RT-NEO methods are used to propagate the dynamics with the assumption of mean-field light-matter interactions. As expected, the two methods yield nearly identical $\mu_{\rm n}^x(t)$ dynamics, although the sc-RT-NEO approach treats the cavity mode classically whereas the mfq-RT-NEO method describes the cavity mode quantum mechanically. The power spectrum of the time-domain signals shown in Fig. \ref{vsc sc mf pc}b yields a pair of LP and UP peaks separated by a 
Rabi splitting of $\Omega_{\rm R}$ = 325 cm\textsuperscript{-1}. Virtually identical  results were obtained with the more computationally efficient mfq-BO-RT-NEO method, which invokes the electronic Born-Oppenheimer approximation during the simulation (see Figure S2). Note that the results in Fig. \ref{vsc sc mf pc} were obtained with methods in which the cavity mode is only coupled directly to the quantum nuclear dipole moment. A comparison of this approach to the total-coupling mfq-RT-NEO method, in which the cavity mode is coupled to both the electronic and quantum nuclear dipole moments, is provided in the SI. \par

%Similar results were obtained with the mfq-BO-RT-NEO method, which invokes the electronic Born-Oppenheimer approximation during the simulation (see Figure S2). This excellent agreement demonstrates the effectiveness of the mfq-BO-RT-NEO method for obtaining coupled molecule-mode real-time dynamics in the VSC regime at greatly reduced computational cost when light-matter interactions are described at the mean-field level. \par

Both the ESC and VSC real-time simulations in Figs. \ref{esc sc mf} and  \ref{vsc sc mf pc} highlight that as long as light-matter interactions are described on a mean-field level, treating the cavity mode classically versus quantum mechanically does not alter the linear-response polariton spectrum. Below, we examine how a fully quantum mechanical treatment of light-matter interactions impacts the time-domain and frequency-domain signals.

\subsection{Strong coupling beyond mean-field light-matter interactions using the full-quantum approach}
\begin{figure}[H]
\centering
\includegraphics[scale=0.4]{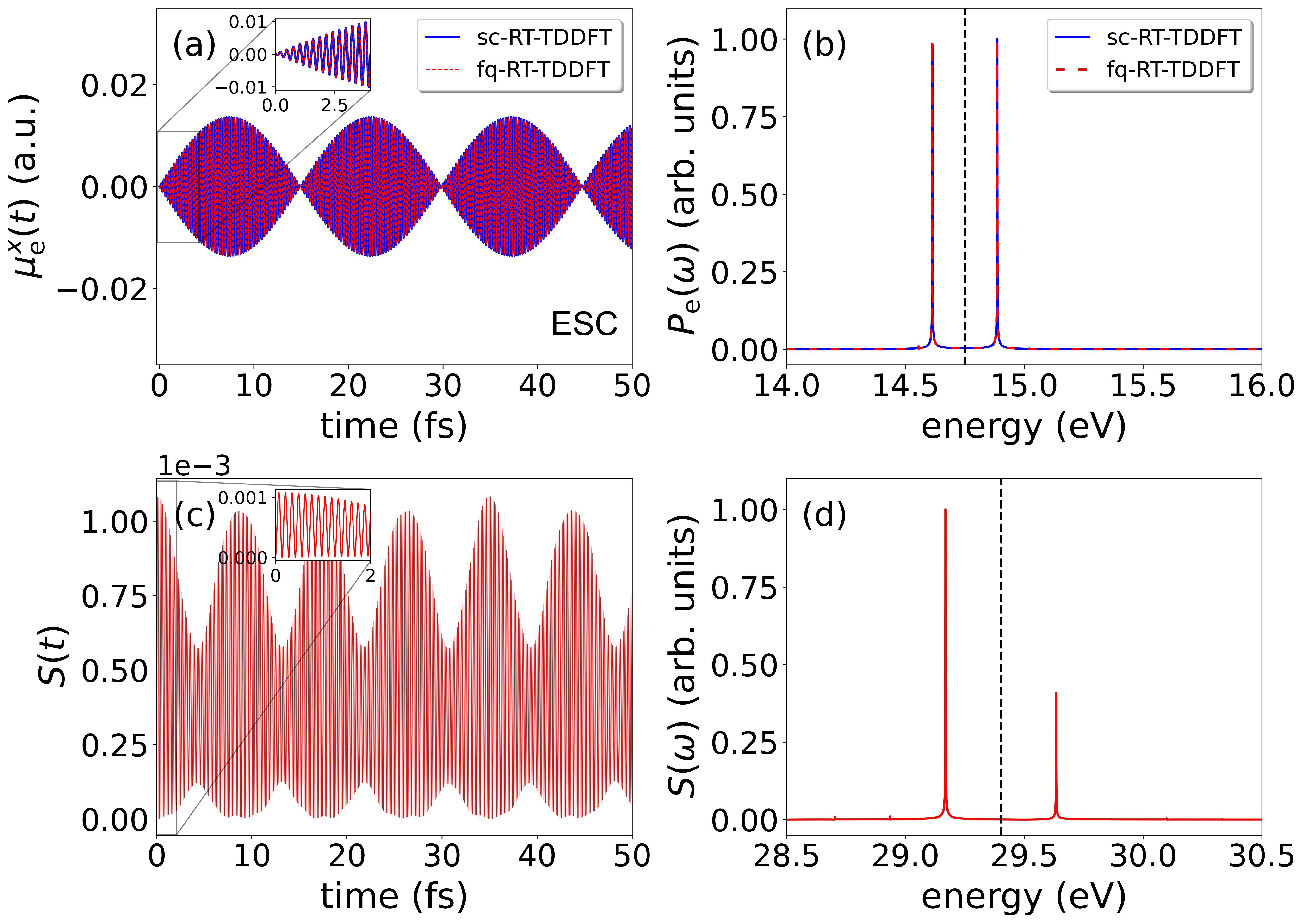}
    \captionsetup{width=1.0\textwidth}
    \caption{(a) sc-RT-TDDFT and fq-RT-TDDFT dynamics of the $x$-component of the H\textsubscript{2} electronic dipole moment $\mu_{\rm e}^x(t)$, with the same settings used as in Fig. 3a. The two signals are virtually indistinguishable. (b) Power spectra $P_{\rm e}(\omega)$ corresponding to the dipole signals in panel a, along with $\omega_{\rm c}$ (vertical black dashed line). The splitting between the two peaks is given by $\Omega_{\rm R}$. (c) von Neumann entropy $S(t)$ computed from the fq-RT-TDDFT calculation according to Eq. 16 using the electronic density matrix $\textbf{P}_{\rm e}(t)$. (d) Fourier transform $S(\omega)$ of the time-domain von Neumann entropy $S(t)$ shown in panel c, along with the fast oscillation frequency $\omega_{\rm N}$ (vertical black dashed line). The splitting between the two major peaks in the spectrum is given by $\Omega_{\rm E}$.
    }
    \label{fq esc}
\end{figure}

We now examine the ESC dynamics for the H\textsubscript{2} molecule studied in Fig. \ref{esc sc mf} by comparing the sc-RT-TDDFT and fq-RT-TDDFT approaches. As introduced in Sec. 2.2, the fully-quantum approach propagates the light-matter dynamics in an extended Hilbert space and models the light-matter coupling by a Kronecker product, thus extending beyond  a mean-field treatment of light-matter interactions. All other simulation details remain identical to those used for the sc-RT-TDDFT calculations.

Fig. \ref{fq esc}a shows the evolution of  $\mu_{\rm e}^x(t)$ predicted by the sc-RT-TDDFT and fq-RT-TDDFT methods after perturbing the cavity mode by a delta pulse at $t = 0$. The two methods yield nearly identical time-domain dynamics (Fig. \ref{fq esc}a) and corresponding power spectra (Fig. \ref{fq esc}b), suggesting that a fully quantum mechanical treatment of light-matter interactions predicts the same linear-response signals under strong coupling as do methods with a mean-field approximation, including the sc-RT-TDDFT method, where the cavity mode is treated classically.

In addition to studying the dipole dynamics, a fully quantum mechanical treatment of light-matter interactions with the fq-RT-TDDFT method permits the study of quantum entanglement between the molecule and cavity mode, which is absent in the mean-field treatment. To investigate this quantum entanglement, we compute the von Neumann entropy $S(t)$ from the electronic density matrix $\textbf{P}_{\rm e}(t)$ in the fq-RT-TDDFT method. This matrix is obtained by taking the partial trace over the field degrees of freedom of the joint density matrix $\textbf{P}_{\rm Fe}(t)$ as given by Eq. \ref{field partial trace}.

As shown in Fig. \ref{fq esc}c, the maximum amplitude of the oscillations of the von Neumann entropy is on the order of $ 10^{-3}$. We can compare this number to the maximum theoretically possible value of $S(t)$ for this calculation, which is given by the natural logarithm of the smaller of the molecular (electronic) basis set size and the cavity mode basis set size. In this case, both basis sets have four functions, so the maximum theoretically possible entropy is $\ln(4) = 1.38$, about three orders of magnitude larger than the largest value of $S(t)$ shown in Fig. \ref{fq esc}c. Our calculations suggest that under single-molecule strong coupling, the molecule-cavity quantum entanglement for this system is very small relative to the maximum theoretically possible value. Due to this small quantum entanglement, it is not surprising that treating light-matter interactions on a fully quantum mechanical level does not alter the linear-response polariton spectrum as compared to lower-level mean-field theories. \par 

Interestingly, despite the magnitude of  $S(t)$, comparison of Figs. \ref{fq esc}a and \ref{fq esc}c reveals that the frequency of the Rabi-like modulating oscillations of $S(t)$ does not match the Rabi oscillation frequency $\Omega_{\rm R}$ in the electronic dipole moment $\mu_{\rm e}^x(t)$. In other words, the quantum entanglement of the molecule and cavity mode subsystems oscillates with a different frequency than the Rabi splitting $\Omega_{\rm R}$, which characterizes the coherent exchange of energy between the two subsystems. We can quantify this difference by considering $S(\omega)$, the Fourier transform  of $S(t)$, as shown in Fig. \ref{fq esc}d. Two peaks appear in $S(\omega)$ in the region of $\sim$ 28 --- 31 eV. The average of the two peaks determines the fast oscillation frequency $\omega_{\rm N}$, which has a numerical value of 29.40 eV and is denoted by the vertical black dashed line. It is interesting to note that $\omega_{\rm N}$ is almost exactly twice $\omega_{\rm c} = 14.75$ eV; this connection will be considered further in future work. Additionally, the two peaks are separated by a splitting $\Omega_{\rm E} = 0.46$ eV, which is roughly twice the Rabi splitting $\Omega_{\rm R} = 0.27$ eV. This rough factor of two can be qualitatively explained in terms  of the molecule-mode entanglement becoming maximized twice in a single precession about the Bloch sphere. This interpretation will also be explored in more detail in future work.

We will hereafter refer to the splitting $\Omega_{\rm E}$ as the ``entanglement Rabi splitting". Since there is no molecule-mode entanglement in the calculations with a mean-field treatment of light-matter interactions (i.e., the sc-RT-TDDFT and mfq-RT-TDDFT approaches), $\Omega_{\rm E}$ is a unique feature of the full-quantum propagation of the light-matter system. It appears that strong coupling can  be characterized by not only the Rabi splitting $\Omega_{\rm R}$ from the linear-response spectrum, but also the entanglement Rabi splitting $\Omega_{\rm E}$ from the von  Neumann entropy.
The maximum observed values of $S(t)$, as well as $\Omega_{\rm R}$, $\Omega_{\rm E}$, and  $\omega_{\rm N}$, calculated with the fq-RT-TDDFT method for H\textsubscript{2} at different coupling strengths are provided in Table S1.

\begin{figure}[H]
\centering
\includegraphics[scale=0.4]{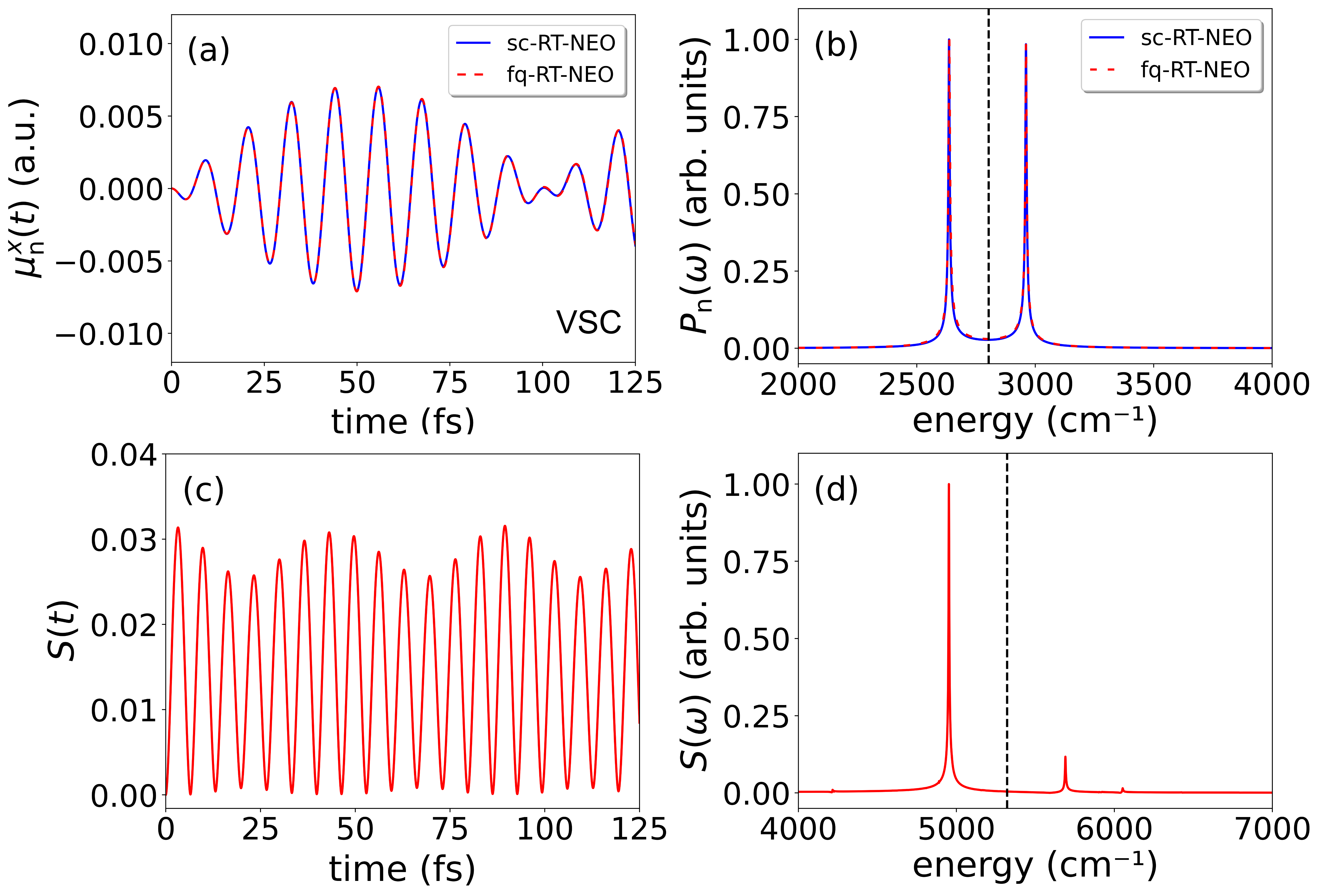}
    \captionsetup{width=1.0\textwidth}
    \caption{(a) sc-RT-NEO and fq-RT-NEO dynamics of the $x$-component of the HCN nuclear dipole moment $\mu_{\rm e}^x(t)$ with the same settings used as in Fig. \ref{vsc sc mf pc}. The two signals are virtually indistinguishable. (b) Power spectra $P_{\rm n}(\omega)$ corresponding to the dipole signals in panel a, along with $\omega_{\rm c}$ (vertical black dashed line). The splitting between the two peaks is given by $\Omega_{\rm R}$. (c) von Neumann entropy $S(t)$ computed using the nuclear density matrix $\textbf{P}_{\rm n}(t)$ from the fq-RT-NEO calculation. (d) Fourier transform $S(\omega)$ of the fq-RT-NEO von Neumann entropy shown in panel c, along with the fast oscillation frequency $\omega_{\rm N}$ (vertical black dashed line). The splitting between the two major peaks is given by $\Omega_{\rm E}$.
    %(i) sc- and fq-BO-RT-NEO dynamics of the $x$-component of the HCN nuclear dipole moment $\mu_e^x(t)$, with the same settings used as Fig. 3c. Both the sc and fq results agree closely with the dynamics shown in panel e. (j) Power spectra $P_n(\omega)$ computed from the dipole signals in panel i. (k) von Neumann entropy $S(t)$ computed from the fq-BO-RT-NEO calculation. (l) Fourier transform $S(\omega)$ of the data in panel k.\
    }
    \label{fig:fq_vsc}
\end{figure}

To examine the universality of our findings under ESC, we now use the fq-RT-NEO approach to investigate the VSC regime for a single HCN molecule coupled to the IR cavity mode, using the same parameters as those used for the sc-RT-NEO and mfq-RT-NEO calculations. 
The fq-RT-NEO and sc-RT-NEO approaches predict virtually identical dipole dynamics (Fig. \ref{fig:fq_vsc}a) and corresponding linear-response polariton spectra (Fig. \ref{fig:fq_vsc}b), both demonstrating a Rabi splitting of $\Omega_{\rm R}$ = 325 cm\textsuperscript{-1}.

The von Neumann entropy $S(t)$ from the fq-RT-NEO calculation is shown in Fig. \ref{fig:fq_vsc}c. Following the fq-RT-TDDFT analysis of H\textsubscript{2} under ESC, we can compare the maximum value of $S(t)$ shown here to the maximum theoretically possible value of $S(t)$ for this calculation. The smaller of the molecular (quantum nuclear) and cavity basis sets is the cavity basis set with four functions, so the theoretical maximum value is once again $\ln(4) = 1.38$. The maximum value of $S(t)$ shown in Fig. \ref{fig:fq_vsc}c is therefore only about 2\% of the maximum value that is theoretically possible for this calculation. More interestingly, by Fourier transforming $S(t)$ to the frequency domain, the $S(\omega)$ spectrum shown in Fig. \ref{fig:fq_vsc}d predicts two peaks with an entanglement Rabi splitting of $\Omega_{\rm E}$ = 735 cm\textsuperscript{-1}, which is nearly twice the value of $\Omega_{\rm R}$, centered around a frequency $\omega_{\rm N} = $ 5321 cm\textsuperscript{-1} (black dashed line) which is nearly twice $\omega_{\rm c}$.  

Similar results were obtained with the fq-BO-RT-NEO method, which invokes the electronic Born-Oppenheimer approximation during the simulation (see Figure S3). This excellent agreement demonstrates the effectiveness of the fq-BO-RT-NEO method for obtaining coupled molecule-mode real-time dynamics in the VSC regime at greatly reduced computational cost. \par

With a full-quantum treatment of light-matter coupling, the ESC and VSC calculations in Figs. \ref{fq esc} and \ref{fig:fq_vsc} suggest that the magnitude of the von Neumann entropy for both systems considered is very small even when the light-matter system is in the strong coupling regime. Despite the small magnitude, the von Neumann entropy exhibits an entanglement Rabi splitting $\Omega_{\rm E}$ under strong coupling, with a frequency of approximately twice that of the polariton Rabi splitting $\Omega_{\rm R}$ for the two systems studied.

\section{Discussion}

In an effort to understand the appearance of the entanglement Rabi splitting $\Omega_{\rm E}$ in the full-quantum dynamics, we investigate the structure of the molecular density matrix using natural orbitals (NOs). While such an investigation is possible for both ESC and VSC calculations, we will focus solely on the ESC case. In our ESC calculations, the small electronic basis set size reduces the number of electronic NOs in the calculation, thus facilitating a succinct analysis. In contrast, for an accurate description of quantum nuclear transitions, the VSC simulation uses a large nuclear basis set, vastly increasing the number of potentially relevant nuclear NOs and making a succinct analysis considerably more difficult. For an H\textsubscript{2} molecule under ESC, the important parameters are $\Omega_{\rm R}$ = 0.27 eV and $\Omega_{\rm E}$ = 0.46 eV, corresponding to oscillation periods of  15.2 fs and 8.9 fs, respectively. 

The results of the analysis that will be presented in the remainder of this section can be summarized as follows. In the full-quantum calculation, the electronic density matrix $\textbf{P}_{\rm e}(t) = \textrm{Tr}_{\rm F}\textbf{P}_{\rm Fe}(t)$ describes an ensemble of two time-dependent pure states. The first state, which dominates the ensemble, exhibits dynamics predominantly on a timescale of $2\pi/\Omega_{\rm R}$ = 15.2 fs. This state bears a strong qualitative resemblance to the single time-dependent pure state that describes the electrons in the semiclassical calculation. The second state, which makes a much smaller contribution to the ensemble, exhibits dynamics on a timescale of $2\pi/\Omega_{\rm E}$ = 8.9 fs. The details of this analysis are provided below.

\subsection{Initial Natural Orbital Occupation Probabilities} 
For this analysis, we compute a basis of NOs from the initial H\textsubscript{2} electronic ground-state density matrix, $\textbf{P}_{\rm e}(0)$. This density matrix is obtained by taking the partial trace over the cavity degrees of freedom of the joint density matrix $\textbf{P}_{\rm Fe}(0)$, which was used as the initial condition of both the semiclassical and full-quantum ESC calculations: $\textbf{P}_{\rm e}(0) = \textrm{Tr}_{\rm F}\textbf{P}_{\rm Fe}(0)$. $\textbf{P}_{\rm e}(0)$ is a representation of the electronic density operator denoted as $\hat{\rho}_{\rm e}(0)$. Diagonalizing this density operator yields a set of NOs $\{\ket{i}\}$ and occupation probabilities $\{p_i(0)\}$ according to the following eigenvalue equation:
\begin{equation}
    \hat{\rho}_e(0)\ket{i} = p_i(0)\ket{i} = \delta_{i1}\ket{i} .
    \label{def_no}
\end{equation}
Here, the kets $\{\ket{i}\}$ are the NOs for the H\textsubscript{2} molecule in the absence of coupling to a cavity mode. We will refer to the states $\{\ket{i}\}$ as ``initial natural orbitals" (INOs) for reasons that will become apparent in the next subsection. In our real-time simulations, only the lowest-energy INO (INO 1) is occupied in the H$_2$ ground state at $t = 0$, so INO 1 has an occupation probability of unity and the others  have an occupation probability of zero, i.e., $p_i(0) = \delta_{i1}$. Therefore, the time-dependent change in INO occupation probability $\langle p_i(t) \rangle$ for the $i$\textsuperscript{th} INO can be defined as
\begin{equation}
    \Delta \langle p_i(t) \rangle \equiv \braket{i|\hat{\rho}_{\rm e}(t)|i} - \braket{i|\hat{\rho}_{\rm e}(0)|i} = \braket{i|\hat{\rho}_{\rm e}(t)|i} - \delta_{i1} ,
    \label{def delta no occ prob}
\end{equation} \par

\begin{figure}[hptb]
\centering
\includegraphics[scale=0.4]{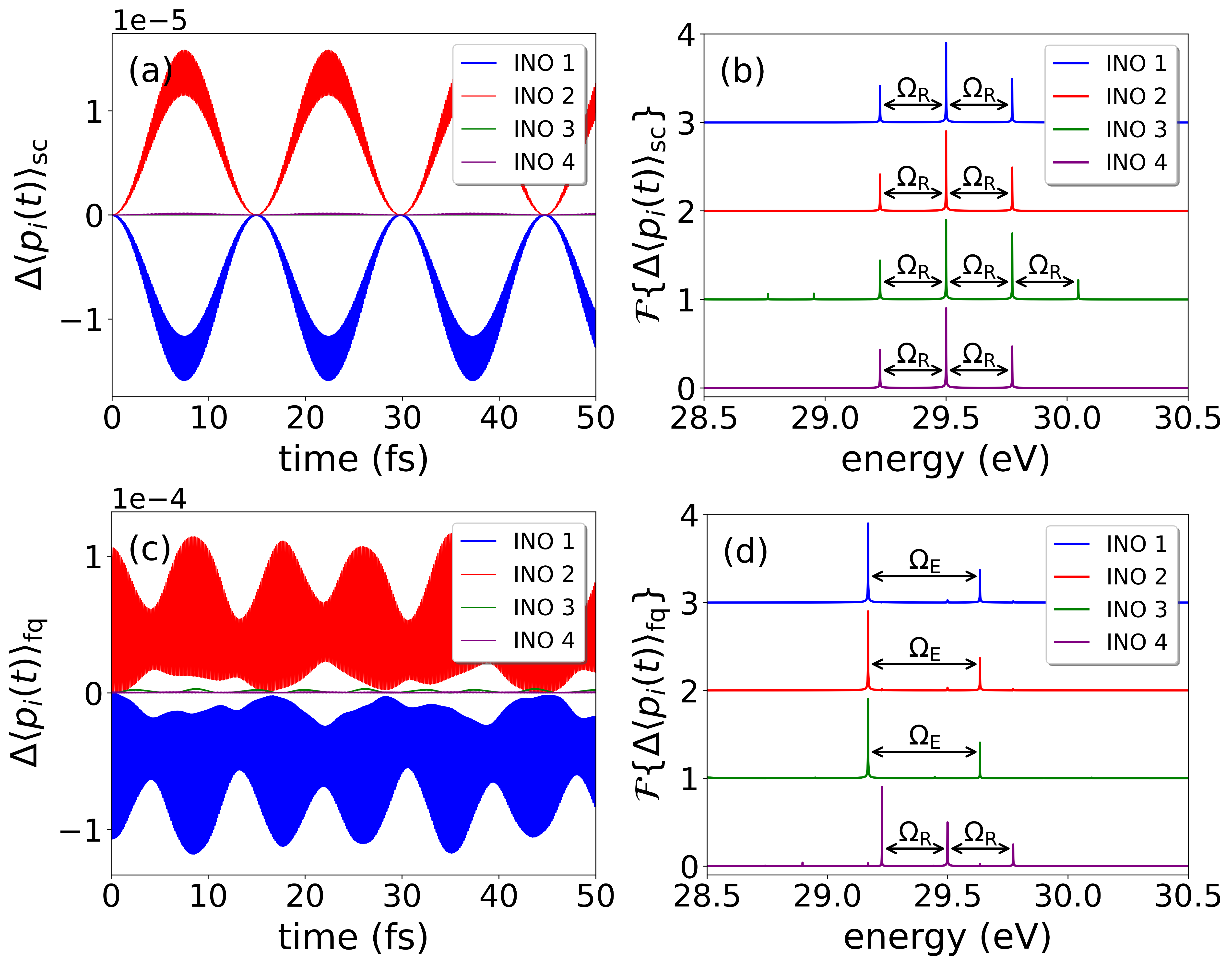}
    \captionsetup{width=1.0\textwidth}
    \caption{(a) Changes in INO occupation probabilities $\Delta \langle p_i(t) \rangle$ computed with sc-RT-TDDFT, oscillating with frequency $\Omega_{\rm R}$. (b) Fourier transforms of the occupation probabilities in panel a. (c) Changes in INO occupation probabilities $\Delta \langle p_i(t) \rangle$ computed with fq-RT-TDDFT. (d) Fourier transforms of the occupation probabilities shown in panel c. Note that each spectrum is scaled independently and shifted along the y-axis for visualization purposes. }
    \label{no occ probs}
\end{figure}

The changes in semiclassical INO occupation probabilities, $\Delta \langle p_i(t) \rangle_{\rm{sc}}$, are shown in Fig. \ref{no occ probs}a. The small amplitude of the oscillations of the probabilities, on the order of  $10^{-5}$, suggests that the ground state INO 1 is the dominant contributor to the electronic density matrix at all times $t$, as the light-matter system is only weakly perturbed at time $t=0$. The dynamics of the occupation probabilities are characterized by $\Omega_{\rm R}$, as indicated by the period of the oscillations, 15 fs, which is consistent with the period of Rabi oscillations, $2\pi/\Omega
_{\rm R} = 15.2$ fs. The corresponding spectra of the INO occupation probabilities are plotted in Fig. \ref{no occ probs}b. In the spectrum of every INO, a characteristic triplet emerges, with both of the two outer peaks split from the central peak by $\Omega_{\rm R}$. The frequency of the middle peak equals the frequency of the fast oscillations, which is 2$\omega_{\rm c} = 2\omega_0$, or twice the cavity mode or molecular transition frequency. The spectrum of INO 3 also shows a side peak to the right of the triplet that is split from the rightmost triplet peak by $\Omega_{\rm R}$. \par

Analogous to Fig. \ref{no occ probs}a, the changes in the full-quantum INO occupation probabilities, $\Delta \langle p_i(t) \rangle_{\rm{fq}}$, are shown in Fig. \ref{no occ probs}c. 
In contrast to the semiclassical INO occupation probabilities, which exhibit dynamics characterized by only $\Omega_{\rm R}$, the full-quantum INO occupation probabilities exhibit dynamics characterized by both $\Omega_{\rm R}$ and $\Omega_{\rm E}$. The slow oscillations of the occupation probabilities of INOs 1 and 2 both have a period of approximately 9 fs, which is consistent with the period of the entanglement Rabi splitting, $2\pi/\Omega_{\rm E} = 8.9$ fs.  In the frequency domain, as shown in Fig. \ref{no occ probs}d,  
the spectra of INOs 1, 2, and 3 exhibit two major peaks separated by $\Omega_{\rm E}$ and centered at $\omega_{\rm N}$ = 29.40 eV,  as shown by $S(\omega)$ in Fig. \ref{no occ probs}d. The spectrum of INO 4 also exhibits this pair of peaks at much weaker intensity. In addition to the pair of peaks from $S(\omega)$, the triplet at $2\omega_0 \pm \Omega_{\rm R} = 29.50 \pm 0.27$ eV seen in the semiclassical INO ocupation probabilities in Fig. \ref{no occ probs}b is observed in the spectrum of INO 4, as well as in the spectra of INOs 1 and 2 at much weaker intensity. Therefore, both $\Omega_{\rm R}$ and $\Omega_{\rm E}$ can be observed in the INO dynamics for the full-quantum calculation. \par

\subsection{Analysis of Time-Dependent Occupation Probabilities and Natural Orbitals}
To better separate the dynamical signals at frequencies $\Omega_{\rm R}$ and $\Omega_{\rm E}$ in the full-quantum calculation, 
we now consider the time-dependent spectral decomposition of  $\textbf{P}_{\rm e}(t)$, which is computed according to $\textbf{P}_{\rm e}(t) = \textrm{Tr}_{\rm F}\textbf{P}_{\rm Fe}(t)$ and is a representation of the operator $\hat{\rho}_{\rm e}(t)$. Previously, we obtained a basis of INOs $\{\ket{i}\}$ by diagonalizing the initial electronic density operator $\hat{\rho}_{\rm e}(0)$. In the same manner, it is also possible to diagonalize the density operator $\hat{\rho}_{\rm e}(t)$ at each timestep, obtaining eigenvalues $\{\Lambda_j(t)\}$ and eigenkets $\{\ket{\Psi_i(t)}\}$: 
\begin{equation}
    \hat{\rho}_{\rm e}(t)\ket{\Psi_i(t)} = \Lambda_i(t)\ket{\Psi_i(t)} .
    \label{time dep diagonalization}
\end{equation}
Here, $\Lambda_i(t)$ represents the $i$\textsuperscript{th} largest eigenvalue of $\hat{\rho}_{\rm e}(t)$ at time $t$, and $\ket{\Psi_i(t)}$ is the corresponding eigenket at time $t$. As  the quantities $\{\Lambda_i(t)\}$ and $\{\ket{\Psi_i(t)}\}$ are evaluated at each timestep by diagonalizing $\hat{\rho}_{\rm e}(t)$, they evolve as continuous functions of time in a similar manner as $\mu_{\rm e}^x(t)$. 
The initial conditions for $\{\Lambda_i(t)\}$ and $\{\ket{\Psi_i(t)}\}$ are $\ket{\Psi_i(0)} = \ket{i}$ and $\Lambda_i(0) = p_i(0) = \delta_{i1}$, signifying that the state $\ket{\Psi_i(t)}$ smoothly evolves from the $i$\textsuperscript{th} INO $\ket{i}$ of the H\textsubscript{2} molecule. We therefore refer to $\ket{\Psi_i(t)}$ as the $i$\textsuperscript{th} time-dependent natural orbital (TDNO) and $\Lambda_i(t)$ as the $i$\textsuperscript{th} time-dependent occupation probability (TDOP). \par

With the TDOPs $\{\Lambda_i(t)\}$ and TDNOs $\{\ket{\Psi_i(t)}\}$,  $\hat{\rho}_{\rm e}(t)$ can be written in terms of its spectral decomposition:
\begin{equation}
    \hat{\rho}_{\rm e}(t) = \sum_i \Lambda_i(t)\ket{\Psi_i(t)}\bra{\Psi_i(t)} . 
    \label{dm spec decomp}
\end{equation}
The physical interpretation of Eq. \ref{dm spec decomp} is that the electronic density operator $\hat{\rho}_{\rm e}(t)$ describes a statistical ensemble of states (TDNOs) $\ket{\Psi_i(t)}$ with probabilities (TDOPs) $\Lambda_i(t)$. As discussed above, this ensemble arises in the full-quantum case as a result of quantum entanglement between the electronic and cavity degrees of freedom; in general, more than one nonzero TDOP $\Lambda_i(t)$ exists at time $t$.  
 
For a better understanding of the time dependence, the TDNOs can be expressed  in terms of the INO basis $\{\ket{i}\}$ defined in Eq. \ref{def_no}:
\begin{equation}
    \ket{\Psi_i(t)} = \sum_j c_{ji}(t)\ket{j}
    \label{tdes expansion} .
\end{equation}
By substituting Eqs. \ref{dm spec decomp} and \ref{tdes expansion} into Eq. \ref{def delta no occ prob}, we obtain an expression for the INO occupation probabilities $\Delta\langle p_i(t) \rangle$ in terms of the TDOPs $\Lambda_j(t)$ and the coefficients $c_{ij}(t)$ of the TDNOs in the INO basis:
\begin{equation}
    \Delta\langle p_i(t) \rangle = \sum_j \Lambda_j(t) |c_{ij}(t)|^2 - \delta_{i1} .
    \label{diagonal elements}
\end{equation}
The sum over the index $j$ runs over the TDNOs. Eq. 38 indicates that the occupation probability of the $i$\textsuperscript{th} INO depends on all of the TDOPs and the coefficients of the $i$\textsuperscript{th} INO in each of the TDNOs. Fig. \ref{no occ probs}d exhibits characteristics of both $\Omega_{\rm R}$ and $\Omega_{\rm E}$ in the full-quantum INO occupation probabilities. Separating the INO occupation probabilities into TDOPs and TDNO coefficients and examining each of these terms individually provides further insights into the roles of $\Omega_{\rm R}$ and $\Omega_{\rm E}$. \par

In the following analysis, we will examine how the reduced electronic density matrix $\textbf{P}_{\rm e}(t) = \textrm{Tr}_{\rm F}\textbf{P}_{\rm Fe}(t)$ carries signals at frequencies $\Omega_{\rm R}$ and $\Omega_{\rm E}$ by studying the TDOPs $\Lambda_i(t)$ and TDNOs $\ket{\Psi_i(t)}$, specifically  the square moduli of the coefficients $|c_{ji}(t)|^2$ in the TDNOs.

\subsubsection{Time-Dependent Occupation Probabilities}

\begin{figure}[H]
\centering
\includegraphics[scale=0.4]{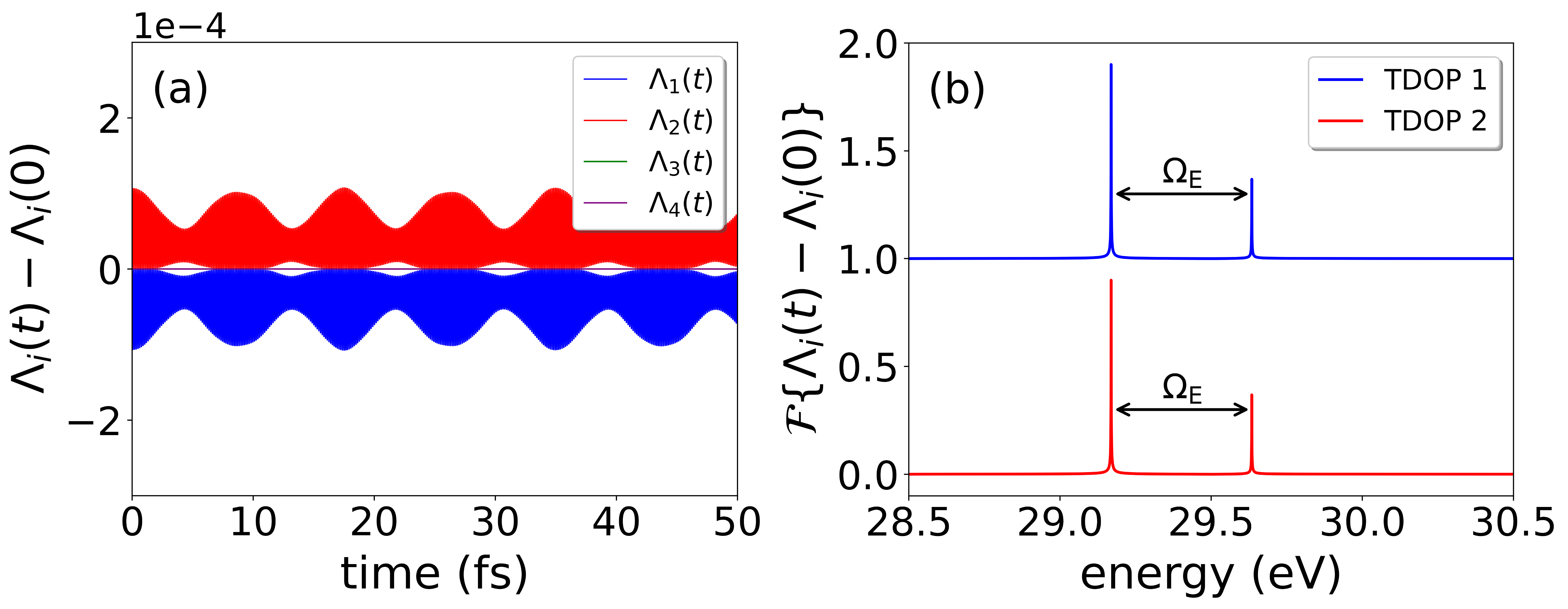}
    \captionsetup{width=1.0\textwidth}
    \caption{(a) Time-dependent occupation probabilities (TDOPs) from the fq-RT-TDDFT calculation. The two largest TDOPs are associated with time-dependent natural orbitals (TDNOs) $\ket{\Psi_1(t)}$ and $\ket{\Psi_2(t)}$. (b) Fourier transforms of TDOPs 1 and 2 in panel a. These peaks are separated by the entanglement Rabi splitting $\Omega_{\rm E}$. Note that each spectrum is scaled independently and shifted along the y-axis for visualization purposes. }
    \label{TDOP}
\end{figure}

Fig. \ref{TDOP}a shows the dynamics of the TDOPs $\Lambda_i(t)$, which oscillate starting from their initial values $\Lambda_i(0) = \delta_{i1}$. Two  TDOPs demonstrate meaningful oscillation amplitudes, denoted as TDOP 1 (blue) and TDOP 2 (red), suggesting significant contributions to the statistical ensemble described by the electronic density operator $\hat{\rho}_{\rm e}(t)$. 
In contrast, the TDOPs 3 (green) and 4 (magenta) are several orders of magnitude smaller than the TDOPs 1 and 2. Due to the weak amplitudes of TDOPs 3 and 4, only the Fourier transforms of the TDOPs 1 and 2 are shown in Fig. \ref{TDOP}b. The peaks in both spectra agree with those in the von Neumann entropy $S(\omega)$ in Fig. \ref{fq esc}d, exhibiting an entanglement Rabi splitting $\Omega_{\rm E}$. 
This agreement is not surprising: inserting  Eq. \ref{dm spec decomp} into Eq. \ref{von neumann} leads to $S(t) = -\sum_i \Lambda_i(t)\ln\Lambda_i(t)$. 

\subsubsection{Time-Dependent Natural Orbital 1, $\ket{\Psi_1(t)}$}
\begin{figure}[H]
\centering
\includegraphics[scale=0.4]{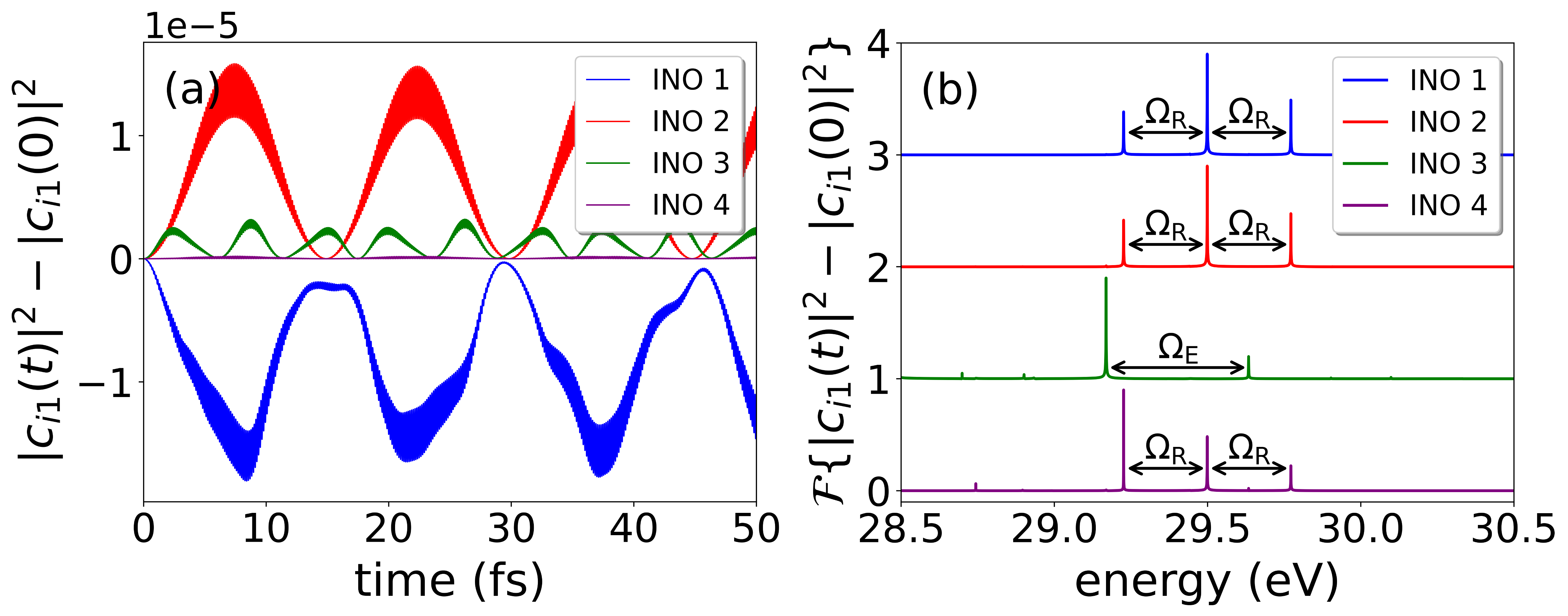}
    \captionsetup{width=1.0\textwidth}
    \caption{(a) Changes in the square moduli of the expansion coefficients of $\ket{\Psi_1(t)}$ in the INO basis from their initial values. The red and blue curves in this figure bears a strong qualitative resemblance to the oscillations of the INO occupation probabilities from sc-RT-TDDFT in Fig. \ref{no occ probs}a. (b) Fourier transforms of the coefficients in panel b, showing the Rabi splitting $\Omega_{\rm R}$. Note that each spectrum is scaled independently and shifted along the y-axis for visualization purposes. }
    \label{TDNO 1}
\end{figure}

We now turn our attention to the TDNOs. Fig. \ref{TDNO 1}a shows the time evolution of the square moduli of the coefficients of $\ket{\Psi_1(t)}$, defined as $|c_{i1}(t)|^2 - |c_{i1}(0)|^2$, with $i=1,2,3,4$ denoting the four INOs defined in Eq. \ref{def_no}. As this TDNO is associated with the TDOP $\Lambda_1(t)$,  
$\ket{\Psi_1(t)}$ is the dominant contributor to the statistical ensemble described by $\hat{\rho}_{\rm e}(t)$. 
Within $\ket{\Psi_1(t)}$, the coefficients of INOs 1, 2, and 4 ($|c_{11}(t)|^2$, $|c_{21}(t)|^2$, and $|c_{41}(t)|^2$)
all demonstrate slow oscillations with a period of 15 fs and are consistent with  $2\pi/\Omega_{\rm R}$.  These oscillations are similar to those in Fig. \ref{no occ probs}a, as further confirmed by comparing the Fourier transform (Fig. \ref{TDNO 1}b) to the Fourier transform of the change in INO occupation probabilities obtained with the semiclassical approach (Fig. \ref{no occ probs}b). The only significant difference between the results in Figs. \ref{TDNO 1}a and \ref{no occ probs}a is observed for INO 3, which demonstrates slow oscillations consistent with $2\pi/\Omega_{\rm E}$, as confirmed by  the appearance of $\Omega_{\rm E}$ associated with the splitting between the peaks in the Fourier transform of INO 3 in Fig. \ref{TDNO 1}b.

To understand the significance of the dynamics shown in Fig. \ref{TDNO 1}, we briefly return to a consideration of the sc-RT-TDDFT approach, which invokes a mean-field approximation for the light-matter interactions, 
so the molecule-mode entanglement is exactly zero. Consequently, its electronic density matrix $\hat{\rho}_{\rm e}(t)$ in the basis of the TDNOs contains only a single time-dependent pure state associated with a single TDNO $\ket{\Psi_{\rm sc}(t)}$:
\begin{equation}
    \hat{\rho}_{\rm e}(t) = \ket{\Psi_{\rm sc}(t)}\bra{\Psi_{\rm sc}(t)} .
    \label{semiclassical density operator}
\end{equation}
In other words, one TDOP is always unity, while all others are always zero. We can then substitute Eq. \ref{semiclassical density operator} and the expansion of $\ket{\Psi_{\rm sc}(t)}$ in the NO basis, $\ket{\Psi_{\rm sc}(t)} = \sum_ic_i(t)\ket{i}$, into Eq. \ref{diagonal elements} and rewrite the semiclassical INO populations $\Delta\langle p_i(t)\rangle_{\rm sc}$ in terms of the coefficients $c_i(t)$:
\begin{equation}
    \Delta\langle p_i(t)\rangle_{\rm sc} = |c_i(t)|^2 - |c_i(0)|^2
    \label{semiclassical no occ prob coeff equivalance}
\end{equation}
Eq. \ref{semiclassical no occ prob coeff equivalance} indicates that the semiclassical INO occupation probabilities shown in Fig. \ref{no occ probs}a are also the square moduli of the coefficients of the pure state $\ket{\Psi_{\rm sc}(t)}$ in the INO basis. Comparison of Fig. \ref{no occ probs}a with the coefficients of $\ket{\Psi_1(t)}$ shown in Fig. \ref{TDNO 1}a suggests that $\ket{\Psi_1(t)}$, the pure state that dominates $\textbf{P}_{\rm e}(t)$ in the full-quantum calculation, strongly resembles the pure state $\ket{\Psi_{\rm sc}(t)}$ that exclusively determines $\textbf{P}_{\rm e}(t)$ in the semiclassical calculation. Therefore, it is not surprising that  $\ket{\Psi_1(t)}$ is mainly responsible for the polariton Rabi splitting $\Omega_{\rm{R}}$.\par

\subsubsection{Time-Dependent Natural Orbital 2, $\ket{\Psi_2(t)}$}
\begin{figure}[H]
\centering
\includegraphics[scale=0.4]{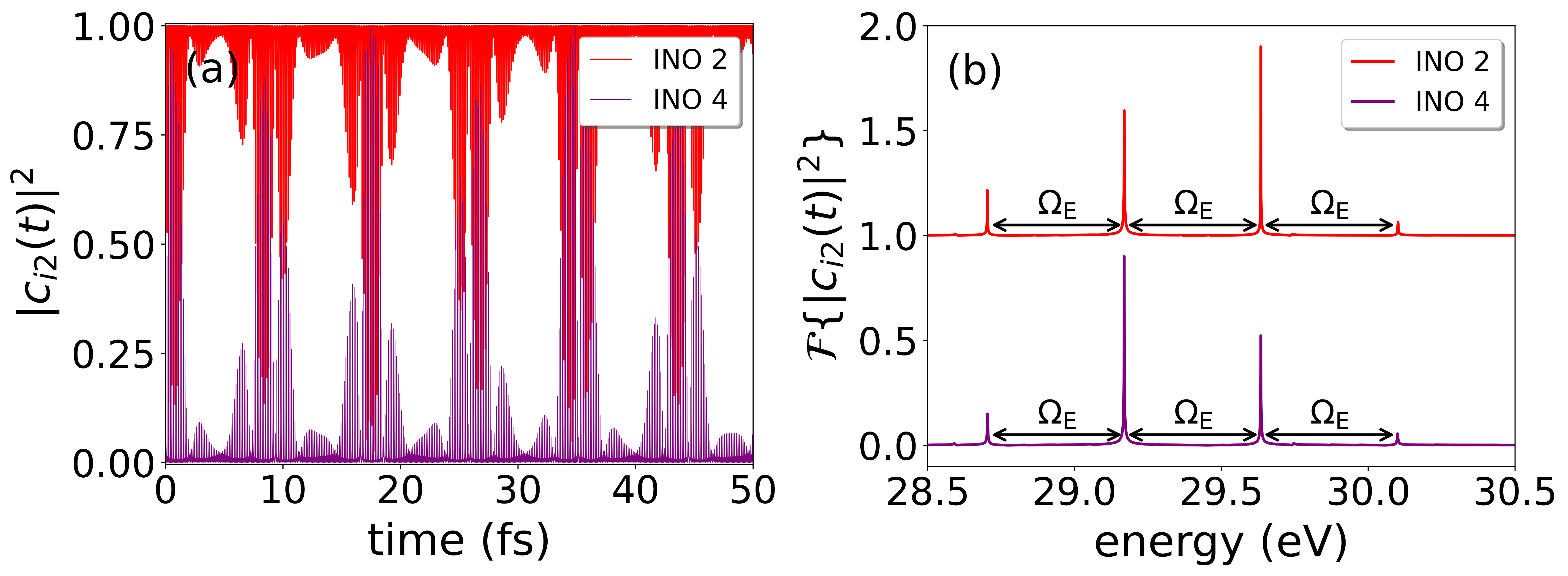}
    \captionsetup{width=1.0\textwidth}
    \caption{(a) Square moduli of the expansion coefficients $\ket{\Psi_2(t)}$ in the NO basis, oscillating with frequency $\Omega_{\rm E}$. (b) Fourier transforms of panel a, showing $\Omega_{\rm E}$. Note that each spectrum is scaled independently and shifted along the y-axis for visualization purposes. }
    \label{TDNO 2}
\end{figure}
Similar to the analysis of $\ket{\Psi_1(t)}$,  Fig. \ref{TDNO 2}a demonstrates the time evolution of the expansion coefficients of $\ket{\Psi_2(t)}$ in the INO basis. As the contributions from INO 1 and 3 are negligible compared to those from INO 2 and 4, the former two contributions are not shown. 
At time $t = 0$, $\ket{\Psi_2(t)} = \ket{2}$, i.e., INO 2, so $|c_{22}(0)|^2 = 1$. At later times, although $|c_{22}(t)|^2$ usually remains close to unity, a near inversion of probability occurs between INOs 2 and 4 with a period of approximately 9 fs, which agrees with the period of the entanglement Rabi splitting $2\pi/\Omega_{\rm E}$. The Fourier transforms of the coefficients of INOs 2 and 4, shown in Fig. \ref{TDNO 2}b, confirm that the frequency of the probability inversion is $\Omega_{\rm E}$. The two main peaks in the spectra of the coefficients of both INOs 2 and 4 are consistent with the peaks in $S(\omega)$ that were used to determine $\Omega_{\rm E}$ in Fig. \ref{fq esc}d. A pair of side peaks appearing in both spectra are also split from the main peaks by $\Omega_{\rm E}$. \par

Our analysis has thus elucidated the manner in which the reduced electronic density matrix $\textbf{P}_{\rm e}(t) = \textrm{Tr}_{\rm F}\textbf{P}_{\rm Fe}(t)$ in the full-quantum calculations carries both frequencies $\Omega_{\rm R}$ and $\Omega_{\rm E}$. In the semiclassical case, $\textbf{P}_{\rm e}(t)$ represents a pure state $\ket{\Psi_{\rm sc}(t)}$ with dynamics characterized exclusively by the polariton Rabi splitting $\Omega_{\rm R}$. In the full-quantum case, $\textbf{P}_{\rm e}(t)$ represents an ensemble dominated by two states, $\ket{\Psi_1(t)}$ and $\ket{\Psi_2(t)}$. The dynamics of $\ket{\Psi_1(t)}$ is predominantly characterized by $\Omega_{\rm R}$ and strongly resembles that of $\ket{\Psi_{\rm sc}(t)}$. In contrast, the dynamics of $\ket{\Psi_2(t)}$ is predominantly characterized by the entanglement Rabi splitting $\Omega_{\rm E}$. \par

\section{Conclusion}
In this paper, we have introduced a hierarchy of methods to simulate first-principles real-time molecule-mode dynamics in the strong-coupling regime. This hierarchy spans semiclassical approaches, where the cavity mode is treated classically, to full-quantum approaches, where the cavity mode is treated quantum mechanically and a joint molecule-mode density matrix is propagated. The full-quantum RT-TDDFT and RT-NEO-TDDFT approaches allow for the inclusion of molecule-mode quantum entanglement effects into real-time dynamics. These approaches were applied to molecular systems in the ESC and VSC regimes, respectively. In both regimes, the Rabi splitting and polariton peak locations are nearly identical for the full-quantum and semiclassical approaches, providing validation for a classical treatment of the cavity modes. \par
Analysis of the molecule-mode entanglement produced in the full-quantum simulations provides additional insights. We find that this entanglement as represented by the von Neumann entropy is several orders of magnitude below its theoretical maximum, suggesting that effects due to molecule-mode entanglement are small, at least in the systems considered in this work. The oscillations of the entanglement, however, are characterized by an entanglement Rabi splitting $\Omega_{\rm E}$ that is different from the Rabi splitting $\Omega_{\rm R}$ computed from the time-dependent dipole moment. More detailed analysis of the electronic density matrix in our calculation of H\textsubscript{2} under ESC reveals that this density matrix describes an ensemble of two pure states. One state behaves qualitatively very similarly to the pure state that occurs when the mode is treated classically in the absence of molecule-mode entanglement. This state is predominantly characterized by $\Omega_{\rm R}$ and is the major contributor to the ensemble. The second state is predominantly characterized by $\Omega_{\rm E}$ and makes a small but still noticeable contribution to the ensemble.\par These findings suggest that although classical electrodynamics is sufficient for the computation of macroscopic observables such as the Rabi splitting, novel physics may be recovered by allowing the molecular subsystem to strongly couple to and entangle with a quantized electromagnetic field. Further analysis of this physics will require the extension of our real-time TDDFT treatment to allow entanglement between a mode and molecular systems with more than two quantized particles entangled to the mode. 
Future work will also aim to address the physical origin of the entanglement Rabi splitting $\Omega_{\rm E}$, the possibility of designing a system for which this splitting may be experimentally observable, and potential effects of the rotating wave approximation on the fast oscillation frequency $\omega_{\rm N}$ of the von Neumann entropy. Consideration of vibronic strong coupling, as in Ref. \citenum{garner_simulation_2025}, also provides an interesting direction for studying light-matter entanglement and could potentially reveal nonlinear optical signatures in spectra. The present work, along with these extensions, will assist in the development of an atomistic-level understanding of the origins of cavity-modified chemistry.

%%%%%%%%%%%%%%%%%%%%%%%%%%%%%%%%%%%%%%%%%%%%%%%%%%%%%%%%%%%%%%%%%%%%%
%% The "Acknowledgement" section can be given in all manuscript
%% classes.  This should be given within the "acknowledgement"
%% environment, which will make the correct section or running title.
%%%%%%%%%%%%%%%%%%%%%%%%%%%%%%%%%%%%%%%%%%%%%%%%%%%%%%%%%%%%%%%%%%%%%
\begin{acknowledgement}

This material is based upon work supported by the Air Force Office of Scientific Research under AFOSR Award No. FA9550-24-1-0347. We thank Marissa Weichman, Jonathan Fetherolf, Scott Garner, Eno Paenurk, Chiara D. Aieta, and Joseph Dickinson for helpful discussions. 

\end{acknowledgement}

%%%%%%%%%%%%%%%%%%%%%%%%%%%%%%%%%%%%%%%%%%%%%%%%%%%%%%%%%%%%%%%%%%%%%
%% The same is true for Supporting Information, which should use the
%% suppinfo environment.
%%%%%%%%%%%%%%%%%%%%%%%%%%%%%%%%%%%%%%%%%%%%%%%%%%%%%%%%%%%%%%%%%%%%%
\begin{suppinfo}
 Discussion and proofs related to properties of the von Neumann entropy; total-coupling mfq-RT-NEO, mfq-BO-RT-NEO, and fq-BO-RT-NEO results on HCN; representative Q-Chem input file and molecular geometry for the reported results.

\end{suppinfo}

%%%%%%%%%%%%%%%%%%%%%%%%%%%%%%%%%%%%%%%%%%%%%%%%%%%%%%%%%%%%%%%%%%%%%
%% The appropriate \bibliography command should be placed here.
%% Notice that the class file automatically sets \bibliographystyle
%% and also names the section correctly.
%%%%%%%%%%%%%%%%%%%%%%%%%%%%%%%%%%%%%%%%%%%%%%%%%%%%%%%%%%%%%%%%%%%%%
\bibliography{Refs}

\end{document}

	%%%%%%%%%% Merge with supplemental materials %%%%%%%%%%
	%%%%%%%%%% Prefix a "S" to all equations, figures, tables and reset the counter %%%%%%%%%%
	\setcounter{equation}{0}
	\setcounter{figure}{0}
	\setcounter{table}{0}
	%\makeatletter
	\renewcommand{\theequation}{S\arabic{equation}}
	\renewcommand{\thefigure}{S\arabic{figure}}
	\renewcommand{\bibnumfmt}[1]{[S#1]}
	\renewcommand{\citenumfont}[1]{S#1}
	\renewcommand{\thepage}{S\arabic{page}}
	
	\maketitle
	
	\newpage

	%%%%%%%%%% Prefix a "S" to all equations, figures, tables and reset the counter %%%%%%%%%%

\tableofcontents

\newpage

\section{Proofs and Discussion of Properties of the von Neumann Entropy}
Here, we provide more detailed proofs and discussion of the properties of the von Neumann entropy $S(t) = -\textrm{Tr}(\hat{\rho}(t)\ln\hat{\rho}(t))$ as enumerated in Section 2.3 of the main text. Throughout this section, we suppress the time dependence of $S(t)$, $\hat{\rho}(t)$, and any terms derived from them, but it should be understood that all quantities below would be time-dependent in a dynamical simulation. \par
\begin{enumerate}
    \item $S$ is nonnegative and minimized to 0 for a pure state, and it is maximized for a random statistical ensemble. If the ensemble has $N$ replicas, then the maximum value of $S$ is $S_{\rm max} = \ln(N)$. $S$ is a sum of terms of the form $-p_i \ln p_i$. This can be seen by inserting the spectral decomposition of $\hat{\rho}$, namely $\hat{\rho} = \sum_i p_i\ket{i}\bra{i}$, into the definition $S = -\textrm{Tr}(\hat{\rho}\ln\hat{\rho})$ to obtain
    \begin{equation}
        S = -\textrm{Tr}\left(\left[\sum_i p_i\ket{i}\bra{i}\right]\ln\left[\sum_j p_j\ket{j}\bra{j}\right]\right) = -\sum_i p_i\ln p_i
        \label{entropy spectral}
    \end{equation}
    Since $p_i \in [0, 1]$ as a probability, $-\ln p_i \geq 0$, and the sum $S$ must therefore be nonnegative since it is a sum of nonnegative terms. For a pure state, the density operator $\hat{\rho}$ reduces to  
    \begin{equation}
        \hat{\rho} = \ket{i}\bra{i}
    \end{equation}
    where $p_i = 1$ is implicit. Applying Eq. \ref{entropy spectral}, we find that $S = 0$. This is the minimum entropy state. The entropy of a random statistical ensemble with $N$ replicas is the entropy of a microcanonical ensemble with $N$ replicas, which is ln($N$). It is physically intuitive that this should be the maximum entropy state; to prove it, we will use Lagrange multipliers. Define the Lagrangian
    \begin{equation}
        \mathcal{L} = - \sum_i^N p_i \ln p_i - \epsilon\left(\sum_j p_j - 1\right)
    \end{equation}
    The stationary condition is 
    \begin{equation}
        \frac{\partial \mathcal{L}}{\partial p_k} = -\ln p_k - 1 - \epsilon = 0
    \end{equation}
    Since this is true for any $k \in \left[1, \textrm{$N$}\right]$, we have 
    \begin{equation}
        -\ln p_1 = -\ln p_2 ... = -\ln p_N \Rightarrow p_1 = p_2 ... = p_N \equiv p
    \end{equation}
    The probabilities must then sum to unity, implying:
    \begin{equation}
        \sum_i p_i = 1 = p \sum_i 1 = Np \Rightarrow p_i = \frac{1}{N}, \; i \in \left[1, N\right] 
    \end{equation}
    The von Neumann entropy is thus
    \begin{equation}
        S = -\sum_i \frac{1}{N} \ln \left(\frac{1}{N}\right) = -\ln\left(\frac{1}{N}\right) = \ln(N)
    \end{equation}
    \item $S$ is invariant under a unitary transformation. This follows immediately from the fact that $S$ is determined by the eigenvalue spectrum of $\hat{\rho}$ (see Eq. \ref{entropy spectral}), which is invariant under a unitary transform, but we make the proof explicit as follows. Let $S$ be the von Neumann entropy of the density operator $\hat{\rho} \equiv \sum_i p_i \ket{i}\bra{i} $ and let $S^\prime$ be the von Neumann entropy of the density operator subjected to a unitary transformation $\hat{U}\hat{\rho}\hat{U}^\dagger$, where $\hat{U}$ is an arbitrary unitary operator. Then
    \begin{equation}
        \begin{split}
            S^\prime &= -\textrm{Tr}\left[\hat{U}\hat{\rho}\hat{U}^\dagger \ln\left(\hat{U}\hat{\rho}\hat{U}^\dagger\right)\right] \\
            &= -\textrm{Tr}\left[\sum_i p_i \hat{U}\ket{i}\bra{i}\hat{U}^\dagger\sum_j \ln p_j \hat{U}\ket{j}\bra{j}\hat{U}^\dagger\right] \\
            &= -\textrm{Tr}\left[\sum_{ij} p_i \ln p_j \hat{U}\ket{i}\bra{j}\hat{U}^\dagger\delta_{ij}\right]\\
            &= -\textrm{Tr}\left[\sum_{i} p_i \ln p_i \hat{U}\ket{i}\bra{j}\hat{U}^\dagger\right]\\
            &= -\sum_{ik} p_i \ln p_i \lvert\braket{k|\hat{U}|i}\rvert^2\\
        \end{split}
    \end{equation}
    The columns of a unitary matrix are orthonormal, so the sum over $k$ is unity and we are left with 
    \begin{equation}
        S^\prime = -\sum_i p_i \ln p_i = S
    \end{equation}
    \item Consider a system AB composed of subsystems A and B and described by a density operator $\hat{\rho}_{\rm AB}(t)$. The von Neumann entropy of the total system AB, denoted as $S_{\rm AB}$, is related to the von Neumann entropies of the subsystems A and B, denoted $S_{\rm A}$ and $S_{\rm B}$, by the inequality 
    \begin{equation}
        S_{\rm AB} \leq S_{\rm A} + S_{\rm B}
        \label{subaddivity}
    \end{equation}
    with equality holding when A and B are independent systems. This property is known as \textit{subaddivity} and can be proved using Klein's inequality. We consider the relative von Neumann entropy $S(\hat{\rho}_1 || \hat{\rho}_2)$ between two density operators $\hat{\rho}_1$ and $\hat{\rho}_2$, which is defined as
    \begin{equation}
        S(\hat{\rho}_1 || \hat{\rho}_2) \equiv \textrm{Tr}(\hat{\rho}_1 \ln \hat{\rho}_1) - \textrm{Tr}(\hat{\rho}_1 \ln \hat{\rho}_2)
    \end{equation}
    Klein's inequality states that $S(\hat{\rho}_1 || \hat{\rho}_2)$ is always nonnegative. This result can then be used to prove Eq. \ref{subaddivity}. The complete proof is given in Ref. \citenum{nielsen_quantum_2009}. 
    
    \item For our bipartite system AB, it is also true that the subsystem entropies $S_{\rm A}$ and $S_{\rm B}$ are equal:
    \begin{equation}
        S_{\rm A} = S_{\rm B}
        \label{subsystem equality}
    \end{equation}
    This result is true regardless of the dimensions of the Hilbert spaces of subsystems A and B. To prove Eq. \ref{subsystem equality}, we need to calculate reduced density matrices for the subsystems A and B and compute the von Neumann entropies for each to show that they are equal. Our proof is a generalization of that given in Ref. \citenum{nielsen_quantum_2009} (which only considers subsystem Hilbert spaces having the same dimension). Let the Hilbert space for subsystem A have dimension $M$ and let the Hilbert space for subsystem B have dimension $N$, where, in general, $M \neq N$. Assuming the total system is in a pure state (as it is throughout our work), we then write the total system's density operator as
    \begin{equation}
        \hat{\rho}_{\rm AB} = \ket{\psi}\bra{\psi}
        \label{total operator}
    \end{equation}
    where the ket $\ket{\psi}$ can be written in general as
    \begin{equation}
        \ket{\psi} = \sum_i^M\sum_j^N \lambda_{ij} \ket{i^{\rm A}}\otimes\ket{j^{\rm B}}
        \label{joint basis expansion}
    \end{equation}
    $\ket{i^{\rm A}}$ and $\ket{j^{\rm B}}$ are base kets for the Hilbert space of subsystems A and B, respectively, so the complete set of states $\ket{i^{\rm A}}\otimes\ket{j^{\rm B}}$ forms a basis for the Hilbert space of the total system AB. The coefficients $\lambda_{ij}$ form an M$\times$N - dimensional complex matrix. We can write this matrix $\boldsymbol{\lambda}$ in terms of its singular value decomposition: 
    \begin{equation}
        \lambda_{ij} = \sum_{k}^M \sum_{l}^N  \boldsymbol{U}_{ik}\boldsymbol{D}_{kl}(\boldsymbol{V}^\dagger)_{lj}
        \label{svd}
    \end{equation}
    Here, $\boldsymbol{U}$ is an $M \times M$ unitary matrix, $\boldsymbol{V}$ is an $N\times N$ unitary matrix, and $\boldsymbol{D}$ is an $M \times N$ rectangular diagonal matrix satisfying R = rank($\boldsymbol{D}$) = min($M$, $N$). The matrix elements of $\boldsymbol{D}$ satisfy
    \begin{equation}
        \boldsymbol{D}_{kl} = 
        \begin{cases}
            k, l \leq R: \; \boldsymbol{D}_{kk}\delta_{kl} \\
            \textrm{else}: \; 0 
        \end{cases}
        \label{D def}
    \end{equation}
    Substituting Eq. \ref{D def} into Equation \ref{svd}, we obtain
    \begin{equation}
        \lambda_{ij} = \sum_{k}^M \sum_{l}^N  \boldsymbol{U}_{ik}\boldsymbol{D}_{kk}\delta_{kl}(\boldsymbol{V}^\dagger)_{lj}
        = \sum_{k}^R \sum_{l}^R  \boldsymbol{U}_{ik}\boldsymbol{D}_{kk}\delta_{kl}(\boldsymbol{V}^\dagger)_{lj}
        = \sum_{k}^R \boldsymbol{U}_{ik}\boldsymbol{D}_{kk}(\boldsymbol{V}^\dagger)_{kj}
        \label{svd 2}
    \end{equation}
    We now substitute Eq. \ref{svd 2} into Eq. \ref{joint basis expansion}:
    \begin{equation}
        \ket{\psi} = \sum_i^M\sum_j^N \sum_{k}^R \boldsymbol{U}_{ik}\boldsymbol{D}_{kk}(\boldsymbol{V}^\dagger)_{kj} \ket{i^{\rm A}}\otimes\ket{j^{\rm B}}
        \label{total state svd}
    \end{equation}
    If we define $\ket{k^{\rm A}} \equiv \sum_i^M\boldsymbol{U}_{ik}\ket{i^{\rm A}}$, $\ket{k^{\rm B}} \equiv \sum_j^N\boldsymbol{V}_{jk}\ket{j^{\rm B}}$, and $\lambda_k \equiv \boldsymbol{D}_{kk}$, then Eq. \ref{total state svd} becomes
    \begin{equation}
        \ket{\psi} = \sum_k^R \lambda_k \ket{k^A} \otimes \ket{k^B}
        \label{schmidt decomp}
    \end{equation}
    Eq. \ref{schmidt decomp} is known as the \textit{Schmidt decomposition} of the pure state $\ket{\psi}$. We have achieved it by changing the bases of the individual subsystems via the unitary matrices $\boldsymbol{U}$ and $\boldsymbol{V}$ so that $\boldsymbol{\lambda}$ is diagonal. Using Eq. \ref{schmidt decomp}, the density operator $\hat{\rho}_{\rm AB}$ can now be written as
    \begin{equation}
        \hat{\rho}_{AB} = \ket{\psi}\bra{\psi} = \sum_{jk} \lambda_j \lambda_k^* \ket{j^{\rm A}}\bra{k^{\rm A}} \otimes \ket{j^{\rm B}}\bra{k^{\rm B}}
    \end{equation}
    We can obtain the spectral representations of each subsystem density matrix by taking the partial trace over the degrees of freedom associated with the other subsystem:
    \begin{equation}
        \hat{\rho}_{\rm A} = \textrm{Tr}_{\rm B}(\hat{\rho}_{\rm AB}) = \sum_k \lvert\lambda_k\rvert^2 \ket{k^{\rm A}}\bra{k^{\rm A}}
    \end{equation}
    \begin{equation}
        \hat{\rho}_{\rm B} = \textrm{Tr}_{\rm A}(\hat{\rho}_{\rm AB}) = \sum_k \lvert\lambda_k\rvert^2 \ket{k^{\rm B}}\bra{k^{\rm B}}
    \end{equation}
    Since the eigenvalue spectra of the two subsystem density operators are equal, it follows immediately by Eq. \ref{entropy spectral} that their von Neumann entropies must be equal. Note that in the notation we have used in this derivation, $\lvert\lambda_i\rvert^2$ here maps to $p_i$ as given in statements 1 and 2.
    \end{enumerate}
\section{Total-Coupling mfq-RT-NEO Results}
\begin{figure}[H]
\centering
\includegraphics[scale=0.4]{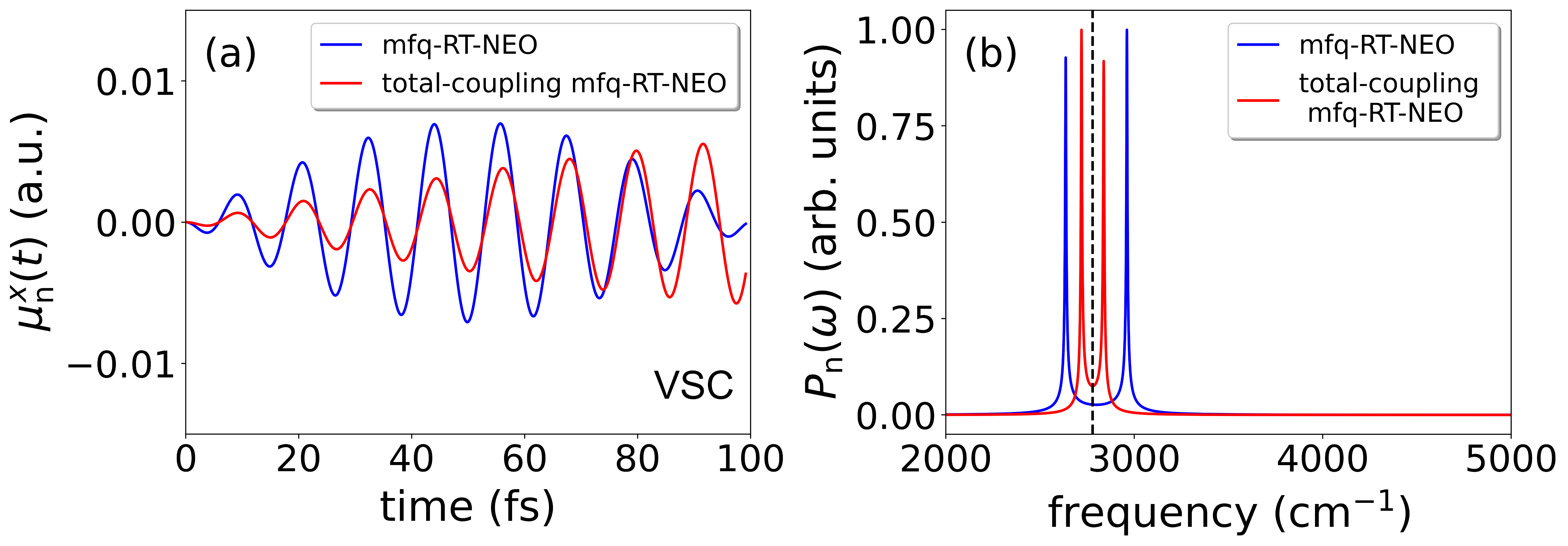}
\caption{(a) mfq-RT-NEO and total-coupling mfq-RT-NEO dynamics of the $x$-component of the HCN nuclear dipole moment $\mu_{\rm n}^x(t)$, coupled to an $x$-polarized cavity mode with frequency $\omega_{\rm c}$ = 2803 cm\textsuperscript{-1} perturbed by a delta pulse at $t = 0$ with coupling strength 8 $\times$ 10\textsuperscript{-4} a.u. (b) Power spectra $P_{\rm n}(\omega)$ corresponding to the dipole signals in panel a, along with $\omega_{\rm c}$ (vertical black dashed line).}
\label{vsc mf pc tc}
\end{figure}

All calculations in the main text assume that the cavity mode is only coupled to the nuclear dipole moment under VSC. In the VSC regime, the coupling between the cavity modes and electrons is expected to be small because of the large difference between electronic and vibrational energy scales. We can test the effects of this assumption by comparing results obtained with the mfq-RT-NEO method to results obtained with what is denoted the total-coupling mfq-RT-NEO method, where the cavity modes are coupled to both the electronic and quantum nuclear dipole moments as follows:
\begin{equation}
    i\hbar \frac{\partial}{\partial t}\textbf{P}_{\rm F}(t) = \left[\textbf{H}_{\rm F} + \sum_{k,\lambda}\varepsilon_{k,\lambda}\textbf{q}_{k,\lambda}\langle\boldsymbol{\mu}_{\rm mol}\rangle_t, \:\textbf{P}_{\rm F}(t)\right] 
    \label{mf rt neo total mode eom}
\end{equation}
\begin{equation}
    i\hbar \frac{\partial}{\partial t}\textbf{P}_{\rm n}(t) = \left[\textbf{F}_{\rm n}(t) + \sum_{k,\lambda}\varepsilon_{k,\lambda}\langle\textbf{q}_{k,\lambda}\rangle_t\boldsymbol{\mu}_{\rm n}, \:\textbf{P}_{\rm n}(t)\right] 
    \label{mf rt neo total nuclear eom}
\end{equation}
\begin{equation}
    i\hbar\frac{\partial}{\partial t}\textbf{P}_{\rm e}(t) = \left[\textbf{F}_{\rm e}(t) +\sum_{k,\lambda}\varepsilon_{k,\lambda}\langle\textbf{q}_{k,\lambda}\rangle_t\boldsymbol{\mu}_{\rm e}, \,\textbf{P}_{\rm e}(t) \right]
    \label{mf rt neo total electronic eom}
\end{equation}
Here, we define $\langle\boldsymbol{\mu}_{\rm mol}\rangle_t \equiv \textrm{Tr}(\textbf{P}_{\rm n}(t)\boldsymbol{\mu}_{\rm n}) - \textrm{Tr}(\textbf{P}_{\rm n}(0)\boldsymbol{\mu}_{\rm n}) + 2\textrm{Tr}(\textbf{P}_{\rm e}(t)\boldsymbol{\mu}_{\rm e}) - 2\textrm{Tr}(\textbf{P}_{\rm e}(0)\boldsymbol{\mu}_{\rm e})$ as the total molecular dipole moment due to quantum nuclei and electrons at time $t$ minus the permanent molecular dipole moment due to quantum nuclei and electrons. \par

We now compare the results from the mfq-RT-NEO and total-coupling mfq-RT-NEO calculations for HCN. 
Fig. \ref{vsc mf pc tc}a shows the $x$-component of the change in the nuclear dipole moment, $\mu_{\rm n}^x(t)$, computed with the mfq-RT-NEO and total-coupling mfq-RT-NEO methods. Using the total-coupling mfq-RT-NEO method decreases the frequency of both the slow and fast oscillations of $\mu_{\rm n}^x(t)$. This behavior is reflected in Fig. \ref{vsc mf pc tc}b, the Fourier transform of the data in Fig. \ref{vsc mf pc tc}a. The frequency of the slow oscillations, given by the Rabi splitting $\Omega_{\rm R}$, decreases from 325 cm\textsuperscript{-1} for the mfq-RT-NEO method to 118 cm\textsuperscript{-1} for the total-coupling mfq-RT-NEO method. \par

The frequency of the fast oscillations is given by the average of the two polariton peak frequencies. For the mfq-RT-NEO method, the two polariton peaks are distributed symmetrically about $\omega_{\rm c}$, so the frequency of the fast oscillations is trivially the cavity mode frequency. For the total-coupling mfq-RT-NEO method, however, the two polariton peaks are distributed asymmetrically about $\omega_{\rm c}$. This asymmetry about $\omega_{\rm c}$ is inconsistent with most experimental results for vibrational polariton peak splittings in molecular systems that remain on the electronic ground state. It also indicates that the fast oscillation frequency no longer matches the cavity mode frequency. Indeed, the value of the fast oscillation frequency in the total-coupling case is 2779 cm\textsuperscript{-1}, which is redshifted from $\omega_{\rm c}$ by 24 cm\textsuperscript{-1}. This redshift can be understood via a second-order perturbative treatment of the three-level model studied by Yang and coworkers\cite{junjie_yang_quantum-electrodynamical_2021, li_electronic_2023}, which demonstrates that coupling between a mode and both a vibrational and electronic transition simultaneously while in the VSC regime induces a redshift in the vibrational polariton peaks. This redshift may be unphysical if it is caused by coupling to higher-energy electronic transitions that are inaccurately described by TDDFT. The effects of the inaccuracies associated with the higher-energy TDDFT electronic states can be mitigated by neglecting the coupling between the cavity modes and the electronic dipole moment. Despite the potential issues of using TDDFT to describe highly excited electronic states, the total-coupling result qualitatively suggests that vibrational polariton peaks may experience a redshift if the molecular system is electronically excited, a prediction that remains open to experimental exploration. \par

These results suggest that more accurate vibrational polaritonic spectra on the electronic ground state may be obtained by assuming that the cavity modes only couple to the quantum nuclear dipole moment. Our conclusion justifies the application of this assumption to the other VSC methods presented in the main text. 

\section{mfq-BO-RT-NEO and fq-BO-RT-NEO Results}
\begin{figure}[H]
\centering
\includegraphics[scale=0.4]{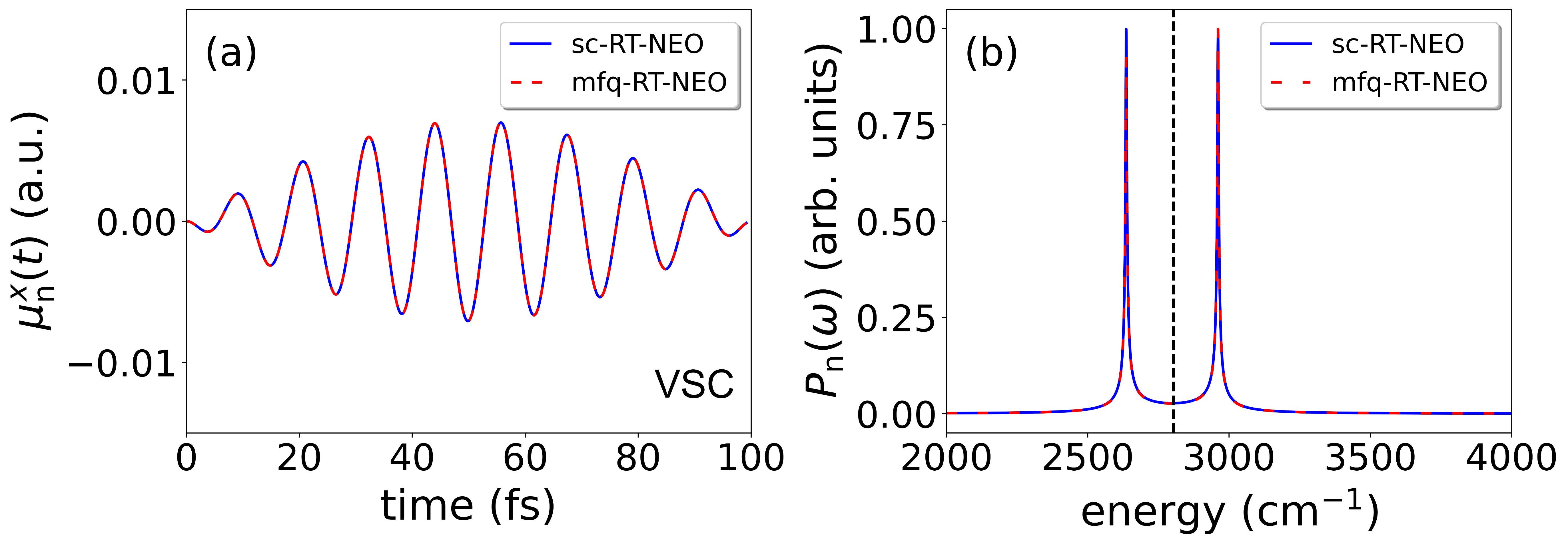}
\caption{(a) sc-BO-RT-NEO and mfq-BO-RT-NEO dynamics of the $x$-component of the HCN nuclear dipole moment $\mu_{\rm n}^x(t)$, coupled to an $x$-polarized cavity mode with frequency $\omega_{\rm c}$ = 2803 cm\textsuperscript{-1} perturbed by a delta pulse at $t = 0$ with coupling strength 8 $\times$ 10\textsuperscript{-4} a.u. (b) Power spectra $P_{\rm n}(\omega)$ corresponding to the dipole signals in panel a, along with $\omega_{\rm c}$ (vertical black dashed line). As expected, the results from the sc-BO-RT-NEO and mfq-BO-RT-NEO methods are indistinguishable.}
\label{vsc mfq bo}
\end{figure}

\begin{figure}[H]
\centering
\includegraphics[scale=0.4]{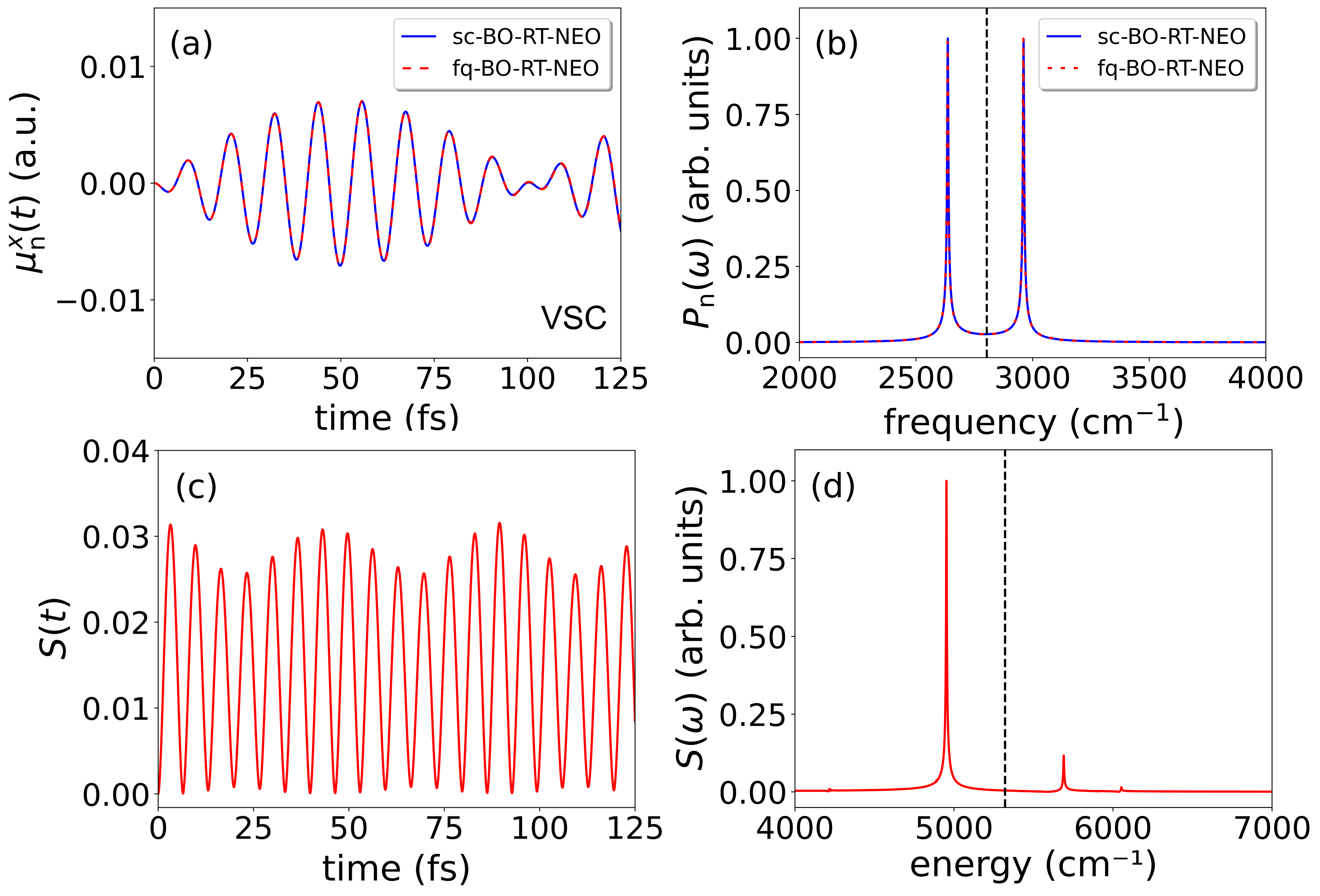}
\caption{(a) sc-BO-RT-NEO and fq-BO-RT-NEO dynamics of the $x$-component of the HCN nuclear dipole moment $\mu_{\rm n}^x(t)$, coupled to an $x$-polarized cavity mode with frequency $\omega_{\rm c}$ = 2803 cm\textsuperscript{-1} perturbed by a delta pulse at $t = 0$ with coupling strength 8 $\times$ 10\textsuperscript{-4} a.u. (b) Power spectra $P_{\rm n}(\omega)$ corresponding to the dipole signals in panel a, along with $\omega_{\rm c}$ (vertical black dashed line). As expected, the results from the sc-BO-RT-NEO and fq-BO-RT-NEO methods are indistinguishable. (c) von Neumann entropy $S(t)$ computed using the nuclear density matrix $\textbf{P}_{\rm n}(t)$ from the fq-BO-RT-NEO calculation. (d) Fourier transform $S(\omega)$ of the fq-BO-RT-NEO von Neumann entropy shown in panel c, along with the fast oscillation frequency $\omega_{\rm N}$ (vertical black dashed line).}
\label{vsc fq bo}
\end{figure}

The same analysis that was carried out in the main paper for the mfq-RT-NEO and fq-RT-NEO simulations can also be carried out for the mfq-BO-RT-NEO and fq-BO-RT-NEO simulations, which invoke the electronic Born-Oppenheimer approximation. As shown by comparing Figs. \ref{vsc mfq bo} and \ref{vsc fq bo} to Figs. 3 and 5 in the main paper, the results computed with the sc-BO-RT-NEO, mfq-BO-RT-NEO, and fq-BO-RT-NEO methods are in very close agreement with the corresponding sc-RT-NEO, mfq-RT-NEO, and fq-RT-NEO results, which do not invoke the electronic Born-Oppenheimer approximation. This agreement demonstrates the effectiveness of the electronic Born-Oppenheimer approximation as a tool for obtaining coupled molecule-mode real-time dynamics in the VSC regime at greatly reduced computational cost. \par

\section{Comparison of Damping Factors in the Pad\'{e} Approximation for mfq-RT-TDDFT Results
}

\begin{figure}[H]
\centering
\includegraphics[scale=0.4]{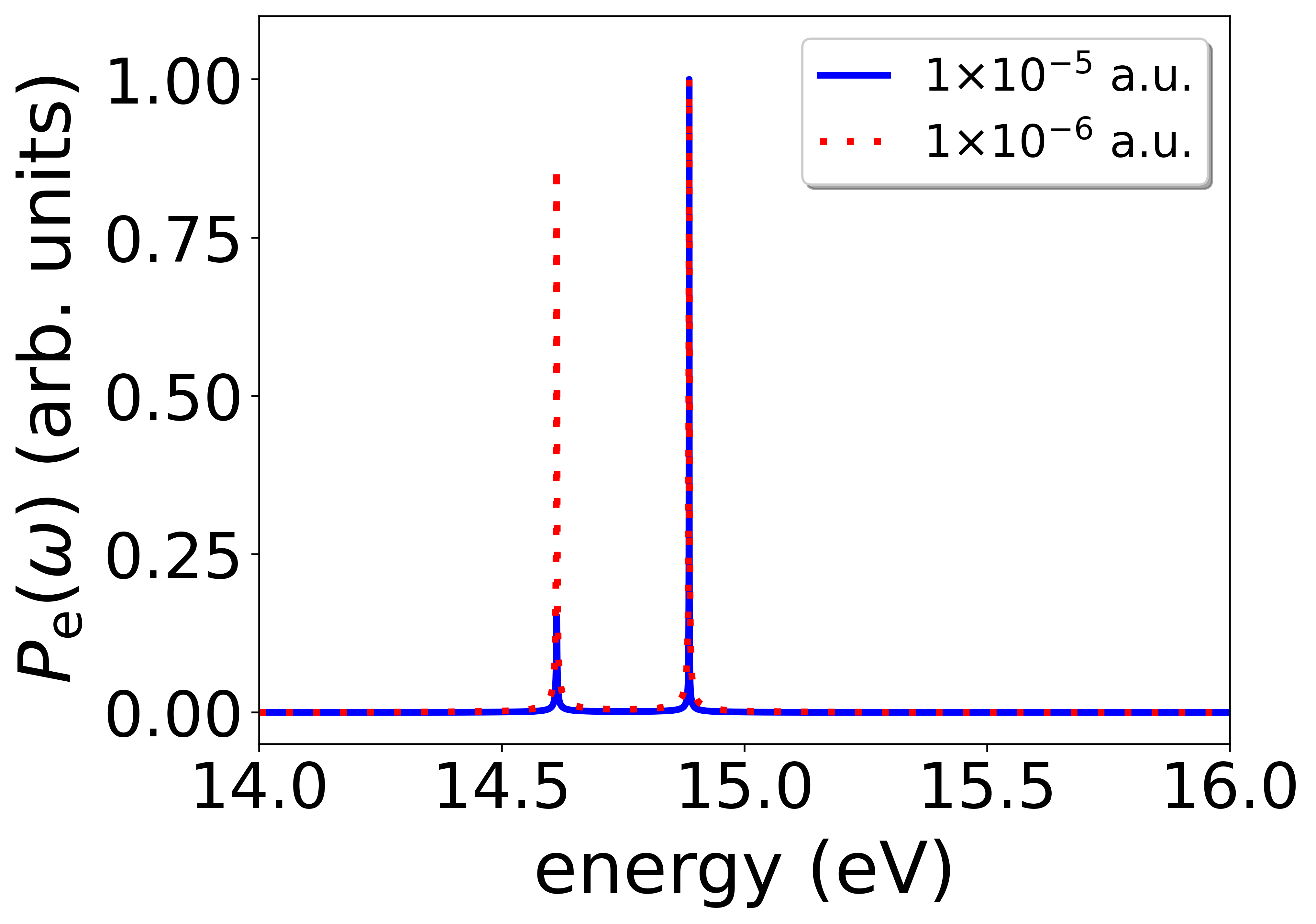}
\caption{Comparison of spectra computed with different damping factors $\gamma$ using the data shown in Fig. 2a of the main text. Decreasing $\gamma$ by an order of magnitude increases the peak intensity from 0.15 to 0.8 without altering the location of the peak.}
\label{pade}
\end{figure}

\section{Rabi Splitting Computed from $\bf q(t)$ with fq-RT-TDDFT}
\begin{figure}[H]
\centering
\includegraphics[scale=0.4]{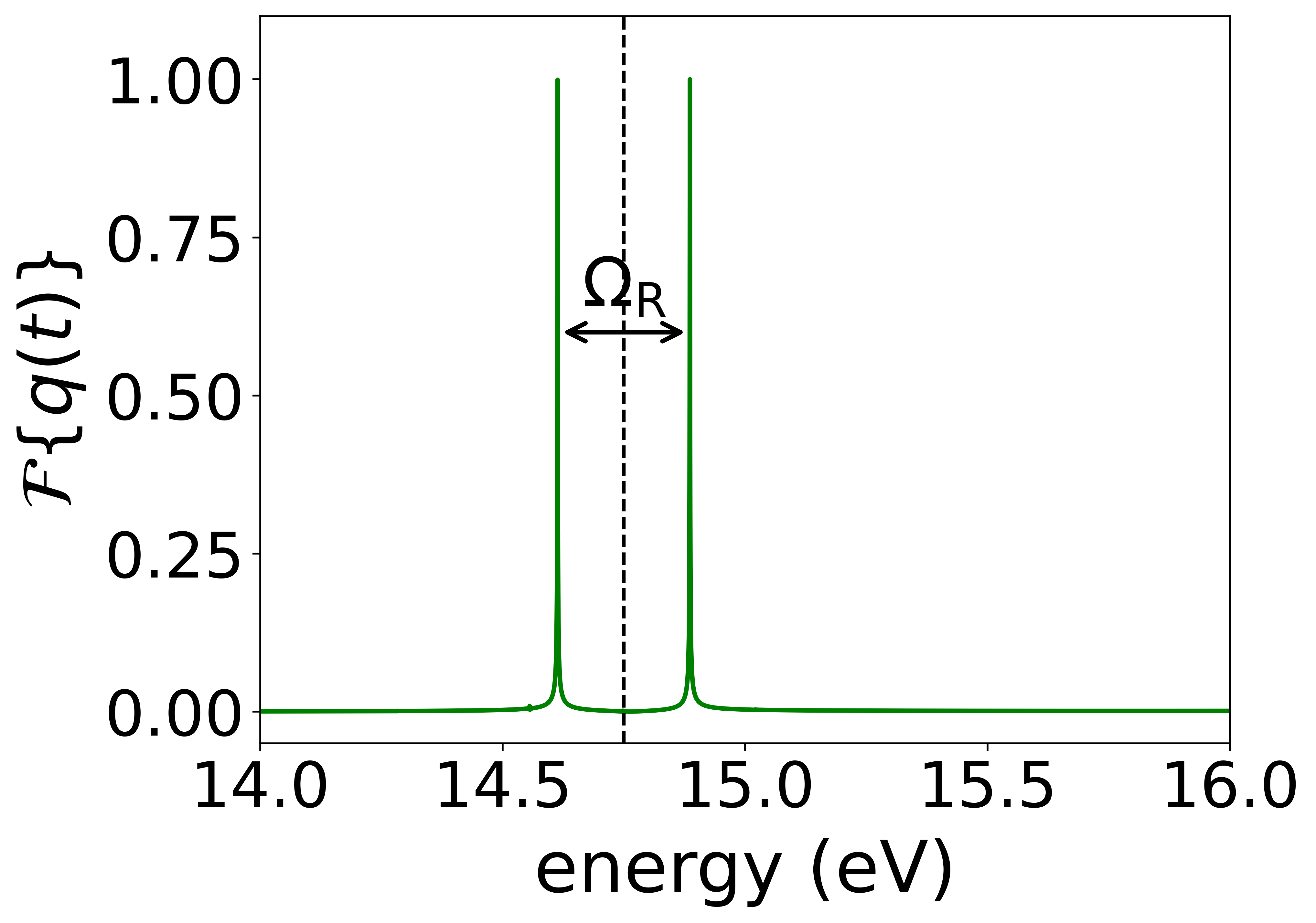}
\caption{Fourier transform of $q(t)$ computed using the fq-RT-TDDFT method. The Rabi splitting $\Omega_{\rm R}$ obtained here is 0.27 eV, which agrees quantitatively with the Rabi splitting computed from the Fourier transform of the dipole signal shown in Fig. 4b.}
\label{q_FT}
\end{figure}

\newpage

\section{Analysis of Entropy, Splittings, and Fast Oscillation Frequency as a Function of Light-Matter 
Coupling Strength}
Table S1 shows the maximum observed entropy max[$S(t)$], Rabi splitting $\Omega_{\rm R}$, entanglement Rabi splitting $\Omega_{\rm E}$, and von Neumann entropy fast oscillation frequency $\omega_{\rm N}$ computed using the fq-RT-TDDFT method for the H\textsubscript{2} system described in the main text at four different light-matter coupling strengths. The coupling strength 0.004 a.u. is the value used to obtain the fq-RT-TDDFT results in the main text. The results in Table S1 show that max[$S(t)$], $\Omega_{\rm R}$, and $\Omega_{\rm E}$ increase with increasing light-matter coupling. The ratio $\Omega_{\rm E}/\Omega_{\rm R}$, however, decreases with increasing coupling strength. As mentioned in the main text, we expect that this ratio should be 2 for a two-level model treatment. For such a model, a single precession about the Bloch sphere should correspond to two passages through the equator of the sphere, where entanglement is maximized. Deviations away from a ratio of 2 reflect the complexity of our first-principles treatment relative to a model system. Finally, the entropy fast oscillation frequency $\omega_{\rm N}$ remains virtually identical for all coupling strengths.

\vspace{10mm} %5mm vertical space

\setcounter{table}{0}
\renewcommand{\thetable}{S\arabic{table}}
\begin{table}[H]
\begin{center}
\caption{Analysis of Entropy, Splittings, and Fast Oscillation Frequency Computed with the fq-RT-TDDFT Method for H$_2$ as a Function of Light-Matter Coupling Strength}
\begin{tabular}{|c c c c c c|} 
 \hline
 Coupling Strength (a.u.) & 
 max[$S(t)$] &
 $\Omega_{\rm R}$ (eV) & $\Omega_{\rm E}$ (eV) & $\Omega_{\rm E}/ \Omega_{\rm R}$ & $\omega_{\rm N}$ (eV)\\ [3.0ex] 
 \hline\hline
 0.001 & 8.6$\times$ 10$^{-5}$ & 0.07 & 0.24 & 3.4 & 29.39 \\ 
 \hline
 0.002 & 3.1$\times$ 10$^{-4}$ & 0.14 & 0.30 & 2.1 & 29.39 \\
 \hline
 0.004 & 1.1$\times$ 10$^{-3}$ & 0.27 & 0.46 & 1.7 & 29.40 \\
 \hline
 0.008 &  3.7$\times$ 10$^{-3}$ & 0.55 & 0.83 & 1.5 & 29.44 \\
 \hline
\end{tabular}
\end{center}
\end{table}

\newpage

\section{Q-Chem Input File for HCN with fq-RT-NEO}
	\RecustomVerbatimCommand{\VerbatimInput}{VerbatimInput}%
	{fontsize=\footnotesize,
		frame=lines,  % top and bottom rule only
		framesep=2em, % separation between frame and text
		label=\fbox{HCN.in\_fq for Fig. 5},
		labelposition=topline,
		commandchars=\|\(\), % escape character and argument delimiters for
		% commands within the verbatim
		commentchar=*        % comment character
	}
	
\begin{Verbatim}
$molecule
0 1
C 0.0 0.0 -0.5026771429
N 0.0 0.0  0.6555628571
H 0.0 0.0 -1.5728771429 
$end

$rem
sym_ignore = 1
input_bohr = false
method = b3lyp
neo = true
neo_epc = epc172
basis = cc-pvdz
SCF_ALGORITHM = GDM
thresh = 14
s2thresh = 12
SCF_CONVERGENCE = 11
NEO_N_SCF_CONVERGENCE = 11
MAX_SCF_CYCLES = 500
NEO_PURECART = 1111
NEO_E_CONV = 12
MEM_TOTAL = 7000
NEO_VPP = 0
NEO_BASIS_LIN_DEP_THRESH = 8 
$end

$neo_basis
H    3
S    1    1.000000
   2.828400 1.0
S    1    1.000000
   4.0 1.0
S    1    1.000000
   5.6569 1.0
S    1    1.000000
   8.0 1.0
S    1    1.000000
   11.3137 1.0
S    1    1.000000
   16.0 1.0
S    1    1.000000
   22.6274 1.0
S    1    1.000000
   32.0 1.0
P    1    1.000000
   2.828400 1.0
P    1    1.000000
   4.0 1.0
P    1    1.000000
   5.6569 1.0
P    1    1.000000
   8.0 1.0
P    1    1.000000
   11.3137 1.0
P    1    1.000000
   16.0 1.0
P    1    1.000000
   22.6274 1.0
P    1    1.000000
   32.0 1.0
D    1    1.000000
   2.828400 1.0
D    1    1.000000
   4.0 1.0
D    1    1.000000
   5.6569 1.0
D    1    1.000000
   8.0 1.0
D    1    1.000000
   11.3137 1.0
D    1    1.000000
   16.0 1.0
D    1    1.000000
   22.6274 1.0
D    1    1.000000
   32.0 1.0
****
$end

$neo_tdks
dt 0.1
photon_type 2
is_coherent true
is_pc true
photon_basis_size 4
scf_e false
maxiter 52500
field_type delta
field_amp 0.3
in_cavity true 0.3475 0 8e-4 1e8
rt_thresh 4
$end
\end{Verbatim}

The above input file generates the data given in Figure 5. The \textit{\$neo\_tdks} section controls the RT-NEO dynamics. ``\textit{photon\_type}" determines the level of theory used and is set to 0 for semiclassical, 1 for mean-field-quantum, and 2 for full-quantum. ``\textit{is\_coherent}" initializes the mode in a coherent state and is set to ``true" for all calculations in this work. ``\textit{scf\_e}" controls the use of the electronic Born-Oppenheimer approximation in RT-NEO calculations. ``\textit{in\_cavity true 0.3475 0 8e-4 1e8}" denotes coupling of the molecule to a single-mode cavity with a mode frequency of 0.3475 a.u., polarization direction $x$ (0 for $x$, 1 for $y$, 2 for $z$), light-matter coupling $\varepsilon = 8 \times 10\textsuperscript{-4}$ a.u., and cavity lifetime $1/\gamma_{\rm c} = 1 \times 10^8$ a.u. The large cavity lifetime indicates that cavity loss in negligible in these simulations. Finally, ``\textit{rt\_thresh}" sets the threshold of the real-time predictor-corrector algorithm to 10\textsuperscript{-4}; this is the \textit{eps} parameter in Algorithm 1 of Ref. \citenum{de_santis_pyberthart_2020}. All other parameters are assumed to be self-explanatory.

\bibliography{SI}